\documentclass[aps,reprint,floatfix,footinbib,notitlepage]{revtex4-2}
\usepackage{natbib}
\usepackage{hyperref}
\usepackage{amssymb}
\usepackage{amsmath}
\usepackage{mathrsfs}
\usepackage{graphicx}
\usepackage{xcolor}
\usepackage{bm}
\usepackage{slashed}

\begin{document}
\title{Interlayer superexchange in bilayer chromium trihalides}

\author{Kok Wee Song}
\affiliation{Department of Physics and Astronomy, University of Exeter, Exeter, Devon EX4 4QL, United Kingdom}

\begin{abstract}
We construct a microscopic model based on superexchange theory for a moir\'e bilayer in chromium trihalides (Cr$X_3$, $X=$Br, I). In particular, we derive analytically the interlayer Heisenberg exchange and the interlayer Dzyaloshinskii-Moriya interaction with arbitrary distances ($\bm{x}$) between spins. Importantly, our model takes into account sliding and twisting geometries in the interlayer $X$-$X$ hopping processes. Our approach can directly access the $\bm{x}$-dependent interlayer exchange without large unit-cell calculations. We argue that deducing interlayer exchange by various sliding bilayers may lead to an incomplete result in a moiré bilayer. Using the \textit{ab initio} tight-binding Hamiltonian, we numerically evaluate the exchange interactions in CrI$_3$. We find that our analytical model agrees with previous comprehensive density functional theory studies. Furthermore, our findings reveal the important role of the correlation effects in the $X$'s $p$ orbitals, which gives rise to a rich interlayer magnetic interaction with remarkable tunability.
\end{abstract}

\date{\today}

\maketitle
\section{Introduction}
Two-dimensional (2D) magnetism in $\mathrm{Cr}X_3$\cite{McGuire:ACScm27(2015),Burch:Nature563(2018),Gibertini:NatNano14(2019),Wang:AnnalPhys532(2020),Yao:Nanotech32(2021),Wang:ACSNano16(2022)} exhibits many fascinating phenomena such as topological magnons\cite{Chen:PRX8(2018),Chen:PRX11(2021),Mook:PRX11(2021),Cai:PRB104(2021),Kvashnin:PRB102(2020),JaeschkeUbiergo:PRB103(2021),McClarty:AnnRewCondMat13(2022)}, moir\'e magnetism\cite{Hejazi:PNAS117(2020),Li:PRB102(2020),Wang:PRL125(2020),Xie:NatPhys18(2021),Xu:NatNanoTech17(2021),Xiao:NanoLett22(2022),Xiao:PRRes3(2021),Akram:NanoLett21(2021),Zheng:AdvFuncMat33(2022),Fumega:2DM10(2023),Yang:NatCompSci3(2023)}, and Kitaev physics\cite{Xu:npjCompMat4(2018),Xu:PRL124(2020),Xu:PRB101(2020),Lee:PRL124(2020),Yadav:arXiv2022}. Interestingly, $\mathrm{Cr}X_3$ may serve as a magnetic building block in forming van der Waal heterostructures\cite{Geim:Nature499(2013),Novoselov:Science353(2016)} for  spintronic applications\cite{Zhong:SciAdv3(2017),Song:Science360(2018),Cardoso:PRL121(2018),Zollner:PRB100(2019),Wang:NatComm9(2018),Kim:NanotLett18(2018),Song:NanoLett19(2019),Rahman:ACSNano15(2021),Heissenbuettel:NanoLett21(2021)}.  Furthermore, the material's unique interlayer exchange coupling is highly tunable leading to intriguing stacking-dependent magnetism\cite{Huang:Nature546(2017),Huang:NatNanotech13(2018),Klein:Science360(2018),Chen:Science366(2019),Kim:PNAS116(2019),Klein:NatPhys15(2019),Li:NatMat18(2019),Guo:ACSNano15(2021),Cheng:NatElec(2023),Xie:NatPhys(2023)}.  In this regard, it has recently attracted much research interest\cite{Sivadas:ACSnano18(2018),Soriano:SSComm299(2019),Jang:PRM3(2019),Jiang:PRB99(2019),Gibertini:JPhysDapp54(2020),Soriano:NanoLett20(2020),Sarkar:PRB103(2021),Wang:JPhysChemC125(2021),Yu:APLett119(2021),Stavric:arXiv(2023)}, particularly in its moir\'e lattice.
Studying these magnetic 2D materials may require comprehensive \textit{ab initio} modeling that can be challenging in a large moir\'e cell. Therefore, an analytical model that can evaluate the spin Hamiltonian accurately is highly desirable. However, this demands an understanding of the interlayer exchange at the microscopic level. Nevertheless, the microscopic origin of the material's interlayer antiferromagnetic (AFM) exchange and its competition with the interlayer ferromagnetic (FM) exchange\cite{Jang:PRM3(2019),Jiang:PRB99(2019)} remains elusive in theory and experiment. 

\begin{figure}
    \centering
    \includegraphics[width=3.35in]{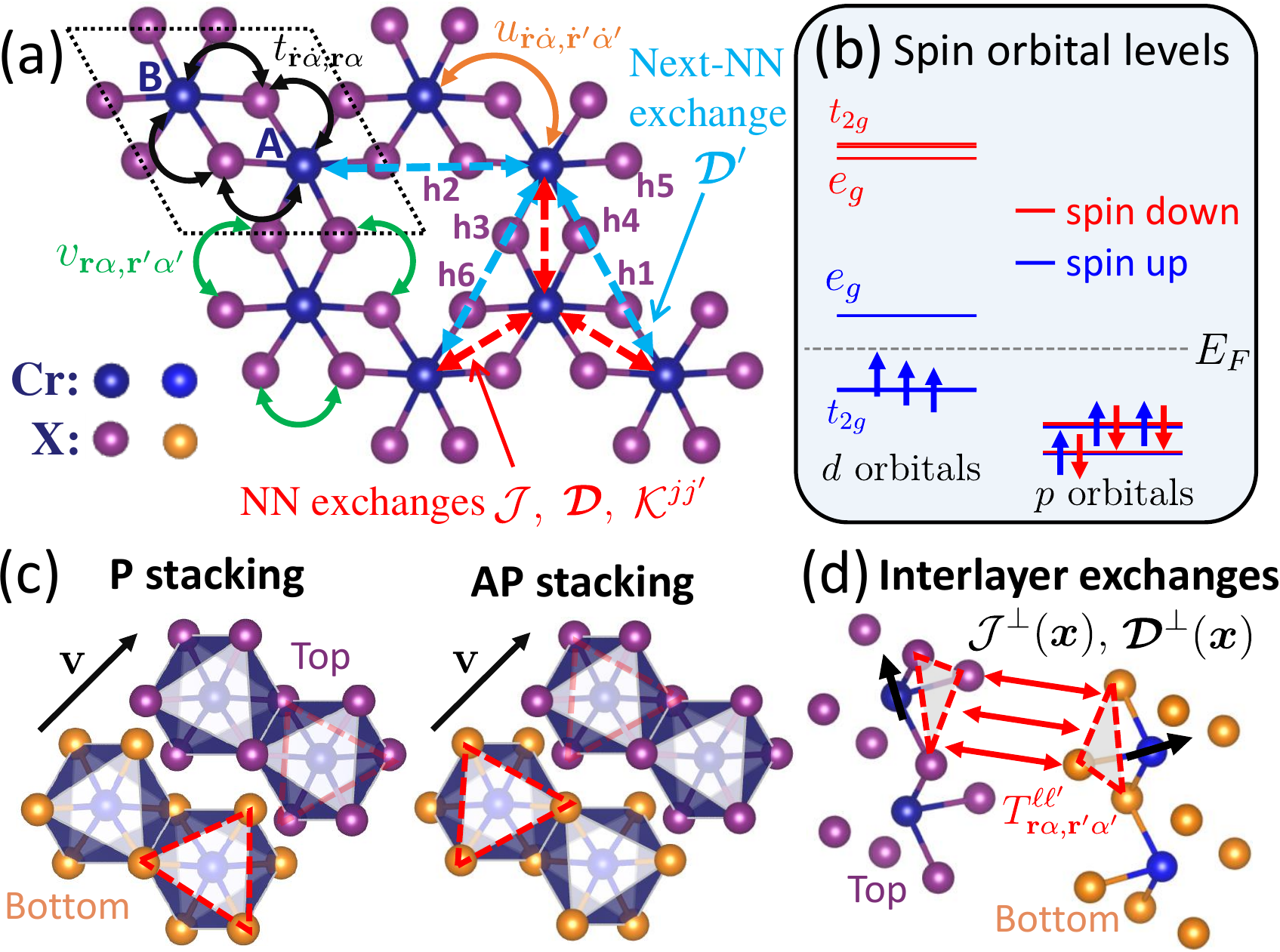}
    \caption{\textbf{Intralayer and interlayer couplings}. (a) Monolayer Cr$X_3$ with hopping constants: $t_{\dot{\mathbf{r}}\dot{\alpha},\mathbf{r}\alpha}$, $u_{\dot{\mathbf{r}}\dot{\alpha},\dot{\mathbf{r}}'\dot{\alpha}'}$, and $v_{\mathbf{r}\alpha,\mathbf{r}'\alpha'}$ and intralayer exchanges, $\mathcal{J}$ (Heisenberg), $\bm{\mathcal{D}}$, $\bm{\mathcal{D}}'$ (DM), and $\mathcal{K}^{jj'}$(symmetric). For each unit cell (dashed parallelogram), it contains two Cr sublattices A and B. (b) The orbital levels for spin-up and spin-down states with Fermi energy $E_F$. (c) The unit cell of a parallel (P) stacking in bilayer Cr$X_3$ (displaced by $\mathbf{v}$ for clarity). Antiparallel (AP) stacking is obtained by rotating the bottom layer by $\pi/3$. (d) The interlayer exchanges $\mathcal{J}^{\perp}$ and $\bm{\mathcal{D}}^{\perp}$ are mediated by the hopping between the $X$ ions located in the two red dashed triangles.}
    \label{fig:Lattice}
\end{figure}

Investigating the interlayer exchange in a Cr$X_3$ moir\'e bilayer is a nontrivial theoretical problem since the exchange coupling mediated by Cr-$X$-$X$-Cr hopping is a complicated process\cite{Sivadas:ACSnano18(2018)} (Fig.\ref{fig:Lattice}d).
Moreover, studying the noncollinear spin order due to the interlayer Dzyaloshinskii-Moriya (DM) interaction induced by spin-orbit coupling (SOC) is also an outstanding question in this moir\'e bilayer. In this paper, we focus on tackling these problems by developing a microscopic model using the superexchange theory\cite{Anderson:PR79(1950),Anderson:PR115(1959),Song:PRB106(2022)}. In our model, we show that the AFM exchange stems from the hopping processes involving the correlated virtual hole pair in the $X$ ion. Also, we will demonstrate that our analytical model yields an accurate spin Hamiltonian for a moir\'e bilayer by performing a small-scale density functional theory (DFT) simulation.

\section{Model} To study the magnetic ground state, we model the bilayer $\mathrm{Cr}X_3$ on-site Hamiltonian\cite{Song:PRB106(2022)} as
\begin{align}
\mathcal{H}_{0}\!=\!\sum_{\ell=1,2}\sum_{\mathbf{R}=\mathbf{r},\dot{\mathbf{r}}}\sum_{\alpha \alpha'}\Big\{[&\epsilon_{\mathbf{R}}^\alpha\delta_{\alpha\alpha'}+U_{\mathbf{R}}^{\alpha \alpha'}(\hat{n}^\ell_{\mathbf{R} \alpha'}-\tfrac{1}{2}\delta_{\alpha \alpha'})]\hat{n}^\ell_{\mathbf{R} \alpha}\notag\\
&-J_{\mathbf{R}}^{\alpha \alpha'}\hat{\mathbf{s}}^\ell_{\mathbf{R}\alpha}\cdot \hat{\mathbf{s}}^\ell_{\mathbf{R}\alpha'}\Big\},\label{eqn:H0}
\end{align}
where, in each layer $\ell$, the $X$'s and Cr's in-plane positions are labeled by $\mathbf{r}$ and $\dot{\mathbf{r}}$. The occupation number and the spin angular momentum operators 
are 
\[(\hat{n}^\ell_{\mathbf{R} \alpha},\hat{\mathbf{s}}^\ell_{\mathbf{R} \alpha})=\frac{1}{2}\begin{cases}
d^{\ell\dagger}_{\dot{\mathbf{r}}\dot{\alpha} \sigma}(\delta_{\sigma \sigma'},\bm{\tau}_{\sigma \sigma'})d^{\ell}_{\dot{\mathbf{r}}\dot{\alpha} \sigma'},&\dot{\alpha}=1\dots5,\\
p^{\ell\dagger}_{\mathbf{r}\alpha  \sigma}(\delta_{\sigma \sigma'},\bm{\tau}_{\sigma \sigma'}) p^{\ell}_{\mathbf{r}\alpha  \sigma'},& \alpha=\tilde{x},\tilde{y},\tilde{z},
\end{cases}
\]
where $p^{\ell\dagger}_{\mathbf{r}\alpha \sigma}$ and $d^{\ell\dagger}_{\dot{\mathbf{r}}\dot{\alpha} \sigma}$ are the creation field operators for $p$ and $d$ orbitals with orbital indces $\alpha$ and $\dot{\alpha}$. Here, $\sigma$ is the spin index which is quantized in the out-of-plane ($z$) direction and $\bm{\tau}=(\tau^x, \tau^y,\tau^z)$ are the Pauli matrices. We note that we use the dotted and undotted indices to explicitly distinguish between $d$ and $p$ orbitals. The model parameters 
$\epsilon^{\alpha}_{\mathbf{R}}$ is the on-site energy, and $U_{\mathbf{R}}^{\alpha \alpha'},J_{\mathbf{R}}^{\alpha \alpha'}>0$ are the on-site Coulomb and Hund interacting constants. The tight-binding (TB) Hamiltonian is
\begin{align}
&\mathcal{H}'=\sum_{\ell \dot{1}1 }(t_{1\dot{1}}p^{\ell\dagger}_{1 \sigma}d^{\ell}_{\dot{1} \sigma}+t_{\dot{1}1}d^{\ell\dagger}_{\dot{1} \sigma}p^{\ell}_{1 \sigma})+\sum_{\ell \dot{1}\dot{2}}u_{\dot{1}\dot{2}}d^{\ell\dagger}_{\dot{1} \sigma}d^{\ell}_{\dot{2} \sigma}\notag\\
&+\sum_{\ell\ell', 12}[(v_{12}\delta_{\sigma \sigma'}\!+\!\bm{\Lambda}_{12}\cdot\bm{\tau}_{\sigma\sigma'})\delta_{\ell \ell'}\!+\!T^{\ell\ell'}_{12}\delta_{\sigma \sigma'}]p^{\ell\dagger}_{1 \sigma}p^{\ell'}_{2\sigma'},\label{eqn:Htb}
\end{align}
where we used the shorthand notation $n\equiv(\mathbf{r}_n\alpha_n)$ and $\dot{n}\equiv(\dot{\mathbf{r}}_n\dot{\alpha}_n)$ to represent the position and the orbital index for $p$ and $d$ orbitals. The intralayer nearest-neighbor (NN) and next-NN hopping constants are $t_{1\dot{1}}$, $u_{\dot{1}\dot{2}}$, and $v_{12}$ [Fig. \ref{fig:Lattice}(a)]. The SOC is written in vector form $\bm{\Lambda}_{12}=-i\lambda\sum_{\alpha}\bm{p}^{\alpha}_{\mathbf{r}_1}\varepsilon_{\alpha\alpha_2\alpha_1}\delta_{\mathbf{r}_2\mathbf{r}_1}$ with coupling strength $\lambda$ and Levi Civita symbol $\varepsilon_{\alpha\alpha_2\alpha_1}$.  We remark that the $p$-orbitals quantization axes $\bm{p}^{\tilde{x},\tilde{y},\tilde{z}}_{\mathbf{r}}$  ($|\bm{p}^{\alpha}_{\mathbf{r}}|=1$) are not arbitrary due to the crystal-field splitting\cite{Song:PRB106(2022)}.

\section{Superexchange theory.}--- The magnetic properties are mostly determined by the low-energy excitations in the Mott's insulating ground state\cite{Anderson:PR79(1950),Anderson:PR115(1959)}. The Mott's state has the many-body wavefunction
\begin{equation}\label{eqn:Mott}
|\tilde{\Psi}\rangle=
\prod_{\ell\dot{\mathbf{r}},\dot{\alpha}\leq3}\chi^{\ell+}_{\dot{\mathbf{r}}\sigma}d^{\ell\dagger}_{\dot{\mathbf{r}}\dot{\alpha}\sigma}\prod_{\ell\mathbf{r}\alpha}p^{\ell\dagger}_{\mathbf{r}\alpha \uparrow}p^{\ell\dagger}_{\mathbf{r}\alpha \downarrow}|0\rangle,
\end{equation}
where $(\chi^{\ell+}_{\dot{\mathbf{r}}\uparrow},\chi^{\ell+}_{\dot{\mathbf{r}}\downarrow})\!=\!
(1\!+\!s^{\ell }_{\dot{\mathbf{r}}z},s^\ell_{\dot{\mathbf{r}}x}\!+\!is^\ell_{\dot{\mathbf{r}}y})/\sqrt{2(1\!+\!s^\ell_{\dot{\mathbf{r}}z})}$ is the local spin wavefunction pointing in $\mathbf{s}^\ell_{\dot{\mathbf{r}}}=(s^\ell_{\dot{\mathbf{r}}x},s^\ell_{\dot{\mathbf{r}}y},s^\ell_{\dot{\mathbf{r}}z})$ with $|\mathbf{s}_{\dot{\mathbf{r}}}^{\ell }|=1$\footnote{The local spin wavefunction with opposite spin direction ($-\mathbf{s}_{\dot{\mathbf{r}}}$) is $\chi^{\ell-}_{\dot{\mathbf{r}}\sigma}=\sum_{\sigma'}i\tau^y_{\sigma \sigma'}\bar{\chi}^{\ell+}_{\dot{\mathbf{r}}\sigma'}$. }.
In the Mott's state, the $p$ orbitals are filled while the $d$ orbitals are half filled in $t_{2g}$ ($\dot{\alpha}=1,2,3$) and empty in $e_g$ [$\dot{\alpha}=4,5$, Fig. \ref{fig:Lattice}(b)]. Because of Hund's interaction, in Eq. \eqref{eqn:Mott}, all the spins at $\dot{\mathbf{r}}$ are deemed to be parallel with $\mathbf{s}^{\ell }_{\dot{\mathbf{r}}}$.

The grand partition function for Mott's state in the interacting picture (treating $\mathcal{H}'$ as a perturbation) is
\begin{align*}
\mathcal{Z}=\sum_{\tilde{\Psi}}\langle\tilde{\Psi}|
&\mathcal{T}e^{-\int_0^\beta d\tau \mathcal{H}'(\tau)}|\tilde{\Psi}\rangle\!\approx \sum_{\text{all possible } \mathbf{s}_{\dot{\mathbf{r}}}}\!\exp \left(-\beta H_{s}\right),
\end{align*}
where $\sum_{\tilde{\Psi}}$ represents the sum in the functional space of the many-body wavefunction $\tilde{\Psi}=\prod_{\ell\dot{\mathbf{r}},\dot{\alpha}\leq3} \chi^{\ell+}_{\dot{\mathbf{r}}\sigma}$. Here, $\mathcal{T}$ is the time-order operator and $\mathcal{H}'(\tau)=\mathrm{e}^{\tau \mathcal{H}_0}\mathcal{H}'\mathrm{e}^{-\tau \mathcal{H}_0}$ with $\tau$ being the imaginary time. Here, $\beta$ is the inverse of temperature. At low temperature, this perturbation expansion\cite{Shankar:RMP66(1994),Song:PRB106(2022)} (seventh order) leads to
\begin{align}
&H_s\!=\!\sum_{\ell}\!\Big[\sum_{\langle\dot{\mathbf{r}} \dot{\mathbf{r}}'\rangle}\!(\mathcal{J}\mathbf{S}^\ell_{\dot{\mathbf{r}} }\cdot\mathbf{S}^\ell_{\dot{\mathbf{r}}'}\!+\!\bm{\mathcal{D}}_{\dot{\mathbf{r}}\dot{\mathbf{r}}'}\!\cdot\mathbf{S}^\ell_{\dot{\mathbf{r}} }\!\!\times\!\mathbf{S}^\ell_{\dot{\mathbf{r}}'}\!+\!\!\sum_{jj'}^{x,y,z}\!\!S^{\ell}_{\dot{\mathbf{r}}j }\mathcal{K}_{\dot{\mathbf{r}}\dot{\mathbf{r}}'}^{jj'}\!S^{\ell}_{\dot{\mathbf{r}}'j'})\!\notag\\
&+\!\!\!\sum_{\langle\langle\dot{\mathbf{r}}\dot{\mathbf{r}}'\rangle\rangle}\!\!\!\bm{\mathcal{D}}_{\dot{\mathbf{r}}\dot{\mathbf{r}}'}'\!\cdot\!\mathbf{S}^\ell_{\dot{\mathbf{r}}}\!\!\times\!\mathbf{S}^\ell_{\dot{\mathbf{r}}'}\!\!+\!\!\sum_{\ell'\dot{\mathbf{r}} \dot{\mathbf{r}}'}^{\ell'\neq\ell}\!(\mathcal{J}^\perp_{\dot{\mathbf{r}} \dot{\mathbf{r}}'}\mathbf{S}^\ell_{\dot{\mathbf{r}} }\!\cdot\!\mathbf{S}^{\ell'}_{\dot{\mathbf{r}}'}\!\!+\!\bm{\mathcal{D}}^\perp_{\dot{\mathbf{r}}\dot{\mathbf{r}}'}\!\cdot\!\mathbf{S}^{\ell}_{\dot{\mathbf{r}} }\!\!\times\!\mathbf{S}^{\ell'}_{\dot{\mathbf{r}}'})\Big]\label{eqn:Hs}
\end{align}
with $\mathbf{S}^\ell_{\dot{\mathbf{r}} }=\frac{3}{2}\mathbf{s}^\ell_{\dot{\mathbf{r}} }$. In the spin Hamiltonian $H_s$, the first three terms are the intralayer NN exchange between AB sublattices [Fig. \ref{fig:Lattice}(a)]. The fourth term is the intralayer NN exchange between AA/BB sublattices. The last two terms are the interlayer Heisenberg exchange $\mathcal{J}^\perp_{\dot{\mathbf{r}}\dot{\mathbf{r}}'}$ and interlayer DM interaction $\bm{\mathcal{D}}^\perp_{\dot{\mathbf{r}}\dot{\mathbf{r}}'}$ which are relevant to the moir\'e magnetism. The derivation of these interlayer exchanges can be found in the Supplemental Material\cite{SM} (SM) and we summarized their analytical expression as follows,
\begin{widetext}
\begin{align}
\mathcal{J}^{ \perp}_{ \dot{\mathbf{r}}_1\dot{\mathbf{r}}_2}\!=&\frac{2^2}{3^2}
\frac{t_{\dot{1} 1}t_{ 4\dot{1}}}{2\mathcal{E}_{\dot{1}1}\mathcal{E}_{\dot{1}4}}\Big\{\frac{P_{\dot{\alpha}_2}T^{\ell_2\ell_1}_{ 34 } t_{ \dot{ 2}3 }
T^{\ell_1\ell_2}_{ 1 2} t_{ 2 \dot{ 2}}}{\mathcal{E}_{\dot{1}2}\mathcal{E}_{\dot{1}3}(\omega_{\dot{1}}-\bar{\omega}_{\dot{2}})}-\frac{t_{\dot{2} 3} T^{\ell_1\ell_2}_{ 12}-\xi t_{\dot{2} 2} T^{\ell_1\ell_2}_{ 13} }{2(\mathcal{E}_{\dot{1}3}\!\!+\!\mathcal{E}_{\dot{2}2}\!\!+\!\Theta^\xi_{23})}
\Big[
\frac{t_{ 2\dot{2}} T^{\ell_2\ell_1}_{ 34}}{\mathcal{E}_{\dot{1}3}^2}+\frac{t_{ 2\dot{2}} T^{\ell_2\ell_1}_{ 34}}{\mathcal{E}_{\dot{2}2}^2}
\Big]\Big\},\quad P_{\dot{\alpha}}=\begin{cases}1,&\dot{\alpha}\leq3\\ 0, &\dot{\alpha}>3 \end{cases};\label{eqn:J_inter}
\\
\bm{\mathcal{D}}^\perp_{\dot{\mathbf{r}}_1\dot{\mathbf{r}}_2}=&i\frac{2^2}{3^2}\frac{t_{\dot{1}1}t_{5\dot{1} }}{\mathcal{E}_{\dot{1}1}\mathcal{E}_{\dot{1}5}}\Big\{\frac{P_{\dot{\alpha}_2} T^{\ell_2\ell_1}_{ 4 5} t_{ \dot{ 2}4}
\Big[T^{\ell_1\ell_2}_{12}\bm{\Lambda}_{23}\!+\!\bm{\Lambda}_{12}T^{\ell_{1}\ell_2}_{23}\Big]t_{3\dot{2}}}{(\omega_{\dot{1}}-\bar{\omega}_{\dot{2}})\mathcal{E}_{\dot{1}2}\mathcal{E}_{\dot{1}3}\mathcal{E}_{\dot{1}4}}
\!+\!\frac{T^{\ell_2\ell_1}_{45}t_{2\dot{2}}\Big(\frac{1}{\mathcal{E}_{\dot{2}2}}\!-\!\frac{1}{\mathcal{E}_{\dot{1}4}}\Big)\!+\!\xi T^{\ell_2\ell_1}_{25}t_{4\dot{2}}\Big(\frac{1}{\mathcal{E}_{\dot{2}4}}\!-\!\frac{1}{\mathcal{E}_{\dot{1}2}}\Big)}{(\mathcal{E}_{\dot{1}2}+\mathcal{E}_{\dot{2}4}+\Theta_{24}^{\xi})}
\Big[\frac{T^{\ell_1\ell_2}_{14}t_{\dot{2}3}\bm{\Lambda}_{32}}{\mathcal{E}_{\dot{2}3}(\mathcal{E}_{\dot{1}1}+\mathcal{E}_{\dot{2}2})}
\notag\\
&
-\frac{T^{\ell_1\ell_2}_{34}t_{\dot{2}2}\bm{\Lambda}_{13}}{\mathcal{E}_{\dot{2}2}(\mathcal{E}_{\dot{1}3}+\mathcal{E}_{\dot{2}2})}-\xi \frac{(T^{\ell_1\ell_2}_{13}\bm{\Lambda}_{32}+\bm{\Lambda}_{13}T^{\ell_1\ell_2}_{32})t_{\dot{2}4}}{\mathcal{E}_{\dot{1}2}\mathcal{E}_{\dot{1}3}}
+\frac{\bm{\Lambda}_{34}}{2}\frac{T^{\ell_1\ell_2}_{12}t_{\dot{2}3}\Big(\frac{1}{\mathcal{E}_{\dot{1}2}}-\frac{1}{\mathcal{E}_{\dot{2}3}}\Big)+\xi T^{\ell_1\ell_2}_{13}t_{\dot{2}2}\Big(\frac{1}{\mathcal{E}_{\dot{2}2}}-\frac{1}{\mathcal{E}_{\dot{1}3}}\Big)}{(\mathcal{E}_{\dot{1}2}+\mathcal{E}_{\dot{2}3}+\Theta_{23}^{-\xi})}
\Big]\Big\},\label{eqn:D_inter}
\end{align}
\end{widetext}
where the sums of all indices are implicitly assumed except $\dot{\mathbf{r}}_{1,2}$. In the above, $\omega_{\dot{n}}$ ($\bar{\omega}_{\dot{n}}$) is the energy for creating a quasielectron (quasihole) in the $d$ orbitals with a spin wave function $\chi^{\ell+}_{\dot{\mathbf{r}}\sigma}$. Namely, the quasielectron (quasihole) state is $|\dot{n}\rangle=\chi^{\ell+}_{\dot{\mathbf{r}}_n\sigma}d^{\ell\dagger}_{\dot{n}\sigma}|\tilde{\Psi}\rangle$ ($|\underline{\dot{n}}\rangle=\bar{\chi}^{\ell+}_{\dot{\mathbf{r}}_n\sigma}d^\ell_{\dot{n}\sigma}|\tilde{\Psi}\rangle$) with $\mathcal{H}_0|\dot{n}\rangle=\omega_{\dot{n}}|\dot{n}\rangle$. Furthermore, $\mathcal{E}_{\dot{n}n}$ is the creation energy of a $d$-electron and $p$-hole pair, $|\dot{n}n\rangle=\bar{\chi}^{\ell+}_{\dot{\mathbf{r}}_n\sigma'} p^\ell_{n \sigma'}|\dot{n}\rangle$, with $\mathcal{H}_0|\dot{n}n\rangle=\mathcal{E}_{\dot{n}n}|\dot{n}n\rangle$. Similarly, the two electron-hole pairs are $|\dot{m}m,\dot{n}n\rangle_{\xi}=\bar{\chi}^{\ell+}_{\dot{\mathbf{r}}_m\sigma'}\bar{\chi}^{\ell+}_{\dot{\mathbf{r}}_n\sigma''} (p^\ell_{m \sigma'}p^\ell_{n \sigma''}+\xi p^\ell_{m \sigma''}p^\ell_{n \sigma'})|\dot{m}\dot{n}\rangle$ where $\xi=\pm1$ for a spin triplet/singlet state with energy $\mathcal{H}_0|\dot{m}m,\dot{n}n\rangle_{\xi}=(\mathcal{E}_{\dot{m}m}+\mathcal{E}_{\dot{n}n}+\Theta_{mn}^\xi)|\dot{m}m,\dot{n}n\rangle_{\xi}$. The spin singlet-triplet energy splitting due to an interaction is given by $\Theta^\xi_{mn}=\frac{1}{2}[U^{\alpha_m \alpha_n}_{\mathbf{r}_m}-(2 \xi+1)(1-\delta_{\alpha_m \alpha_n})J^{\alpha_m \alpha_n}_{\mathbf{r}_m}]\delta_{\mathbf{r}_m \mathbf{r}_n}$. For simplicity, in Eqs.\eqref{eqn:J_inter} and \eqref{eqn:D_inter}, the high-energy virtual states in the $d$ orbitals\cite{SM} [spin-down states in Fig. \ref{fig:Lattice}(b)] are projected out\cite{Song:PRB106(2022)}. For calculating the exchange coupling of any two spins, we consider only the first nine shortest paths for the interlayer $X$-$X$ hopping which is depicted in Fig. \ref{fig:Lattice}(d).

\begin{figure}
    \centering
    \includegraphics[width=3.3in]{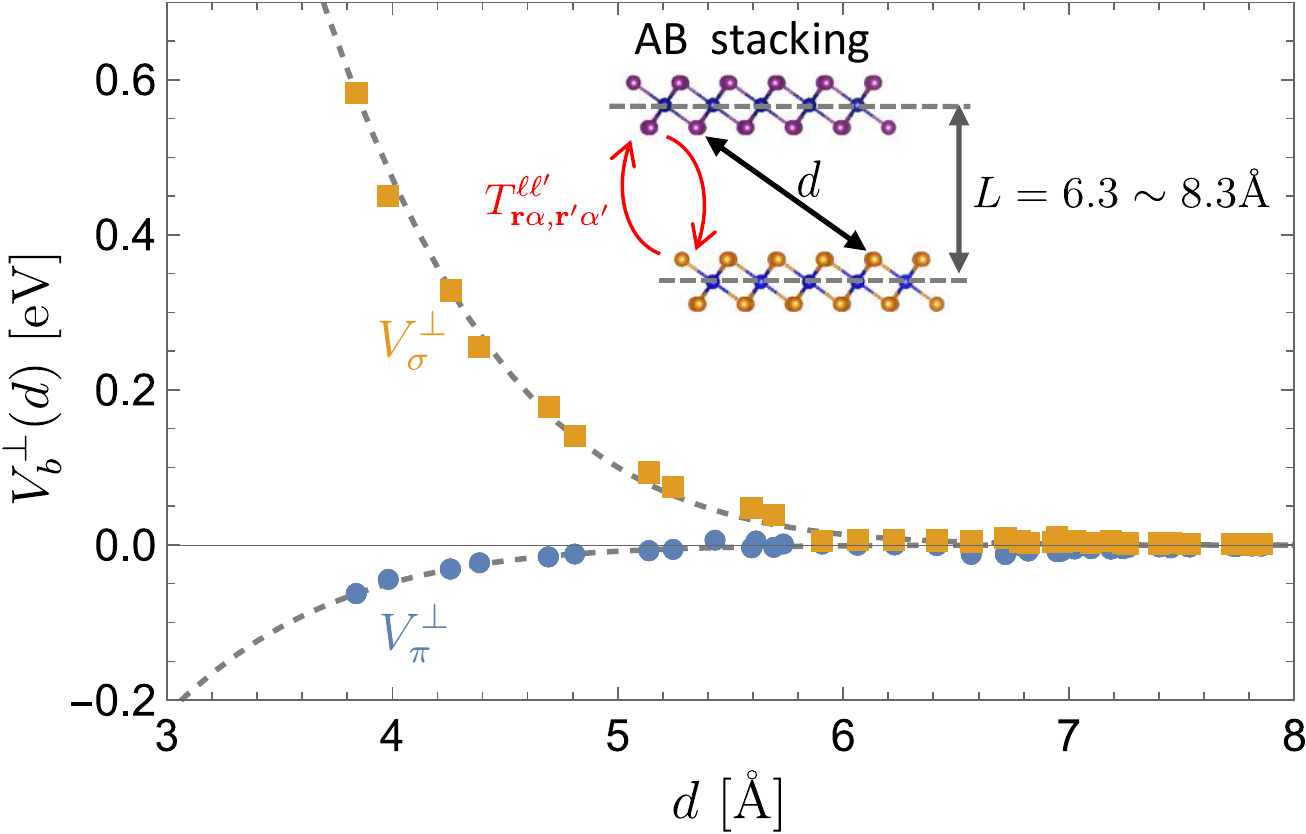}
    \caption{\textbf{Koster-Slater parameters}. The dashed curves are fitted by using Eq. \eqref{eqn:Tinter}. The KS integral is obtained by the DFT calculation with the AB-stacking bilayer (P stacking). The interlayer distance increases in the steps of $0.5$ \AA.}
    \label{fig:KS}
\end{figure}

\section{Interlayer exchange in CrI$_3$}
To estimate the interlayer exchange coupling, we model the interlayer hopping as 
\begin{equation}\label{eqn:Tinter}
T^{\ell\ell'}_{\mathbf{r}\alpha,\mathbf{r}'\alpha'}\!=\!\delta_{\alpha \alpha'}V_\pi^{\perp}\!+\!(V_\sigma^{\perp}\!-\!V_\pi^{\perp})\frac{ \bm{p}^\alpha_{\mathbf{r}}\!\cdot\bm{d}  \bm{p}^{\alpha'}_{\mathbf{r}'}\!\cdot\bm{d}}{d^2},
\end{equation}
where $\bm{d}=\mathbf{r}+\mathbf{h}_{\ell}-(\mathbf{r}'+\mathbf{h}_{\ell'})$ is the displacement between two $X$ ions with
$\mathbf{h}_{1,2}=\pm \frac{1}{2}(0,0,L)$. Here, $L$ is the interlayer distance between Cr planes (Fig.\ref{fig:KS}). The Slater-Koster (SK) overlap integral is parametrized\cite{Guinea:PRB99(2019)} by a $\bm{d}$-dependent function as
\begin{equation}\label{eqn:Vb}
V_{b}^{\perp}=\nu_{b}\exp[-(d/R_b)^{\xi_b}], \quad b=\pi,\sigma.
\end{equation} 
To find the SK parameters $\nu_b$, $R_b$, and $\xi_b$, we compute $T^{\ell\ell'}_{\mathbf{r}\alpha,\mathbf{r}'\alpha'}$ by constructing an \textit{ab initio} TB Hamiltionian using QUANTUM ESPRESSO\cite{Giannozzi:JPhysCondMat29(2017)}, WANNIER90\cite{Marzari:RMP84(2012),Pizzi:JPhysConfMat32(2020)}, and the pseudopotential from the standard solid-state pseudopotential efficiency library\cite{Lejaeghere:Science351(2016),Prandini:njpCompMat4(2018)} (see SM\cite{SM}). We then perform the fitting (Fig.\ref{fig:KS}) for the SK parameters\cite{Fang:PRB92(2015)} by using $T^{\ell\ell'}_{\mathbf{r}\alpha,\mathbf{r}'\alpha'}$ with various interlayer distances ($L=6.3$--$8.3$\AA) and the result is summarized in Table \ref{tbl:Vsp}.

\begin{table}
\caption{\label{tbl:Vsp}Koster-slater integral $V_{b}^{\perp}=\nu_{b}\exp[-(d/R_b)^{\xi_b}]$ for interlayer hopping.}
\begin{ruledtabular}
\begin{tabular}{lccc}
$b$&$\nu_{b}$ & $R_{b}$&$\xi_{b}$\\
\hline
$\sigma$ & 5.3547&2.6794&2.2148\\
$\pi$ & -2.3698&1.7993&1.7021\\
\end{tabular}
\end{ruledtabular}
\end{table}
We then proceed to obtain the TB constants and $\bm{p}^\alpha_{\mathbf{r}}$ in $\mathcal{H}'$ by extracting them from the spin-up \textit{ab initio} TB Hamiltonian. Using the spin-down TB constants only leads to minor modifications\cite{Song:PRB106(2022)}. However, we note that the correlation energy $\Theta^{\xi}_{\dot{n}n}$ cannot be obtained by the one-particle Kohn-Sham spectrum. Therefore, the interactions between $p$ holes, $U_{\mathbf{r}_n}^{\alpha\alpha'}$ and $J_{\mathbf{r}_n}^{\alpha\alpha'}$, remain free parameters in our model. Here, we use $U_{\mathbf{r}_n}^{\alpha\alpha'}=1.2$ eV and $J_{\mathbf{r}_n}^{\alpha\alpha'}=0.5$eV from Ref.\cite{Song:PRB106(2022)}. Once all the model parameters are determined, we can calculate the superexchange in Eqs. \eqref{eqn:J_inter} and \eqref{eqn:D_inter}. But, the quasiparticle energies between the spin-up $t_{2g}$ (valence) and $e_g$ (conduction) bands [Fig. \ref{fig:Lattice}(b)] in the \textit{ab initio} TB Hamiltonian are inaccurate due to the band-gap underestimation in DFT\cite{Song:PRB106(2022)}. To correct this, it requires performing a $GW$ calculation\cite{Wu:NatComm10(2019),Acharya:PRB104(2021)} which is beyond the scope of this paper. Instead, we employ the ``scissor" correction\cite{Fiorentini:PRB51(1995),Johnson:PRB58(1998),Bernstein:PRB66(2002),Magorrian:PRB94(2016)} by rigidly shifting the conduction $e_g$ bands by an additional $1.2$ eV \cite{SM} above the valence $t_{2g}$ band to match the intralayer exchange\cite{Zhang:JMatChemC3(2015),Lado:2DMat4(2017),Besbes:PRB99(2019),Torelli:2DMat6(2019),Wu:PhysChemChemPhys21(2019),Kashin:2DMat7(2020),Stavropoulos:PRR3(2021)}, $\mathcal{J}\sim4$ meV.

\begin{figure}
\centering
    \includegraphics[width=3.35in]{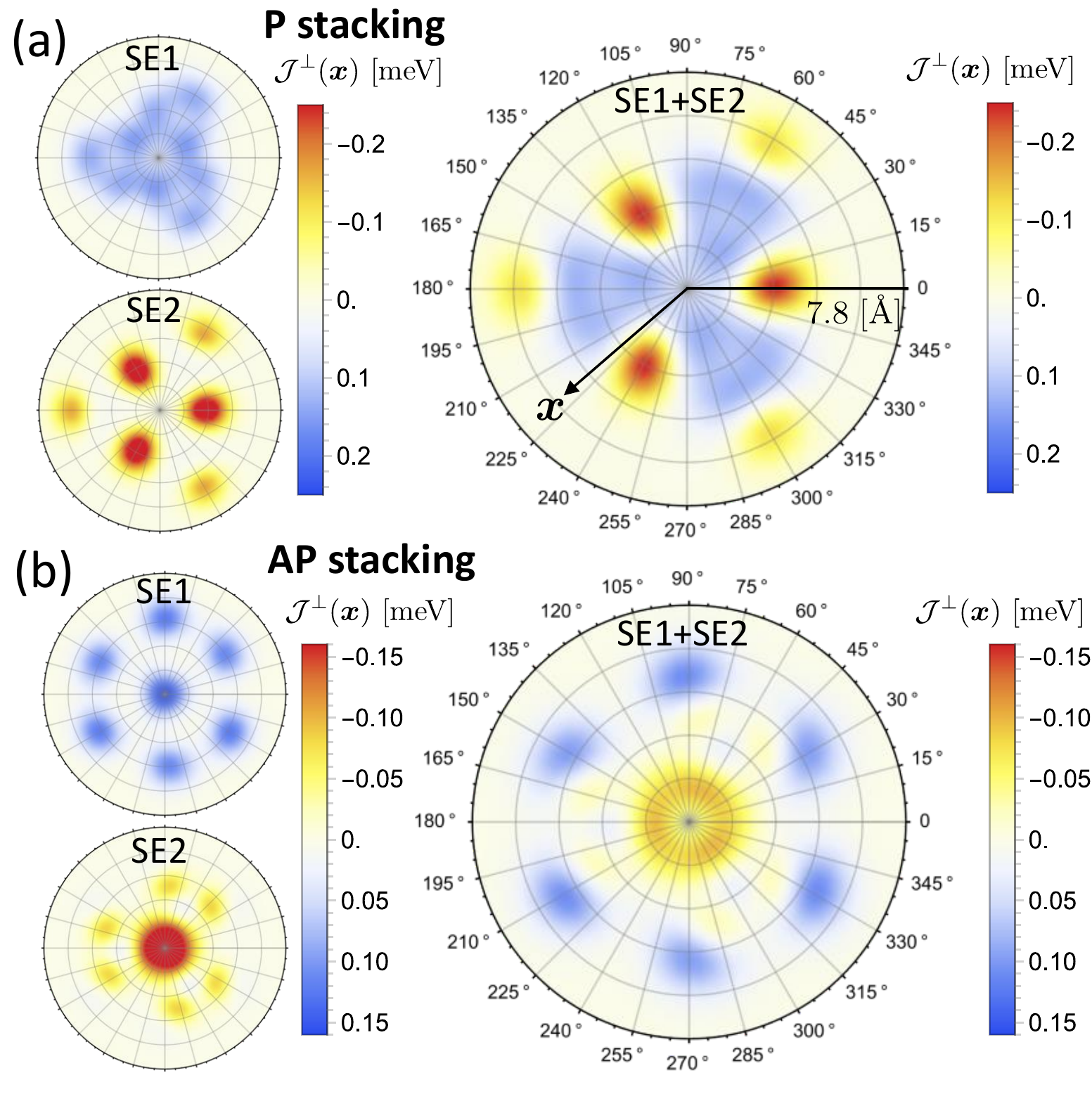}
    \caption{\textbf{Interlayer exchange coupling}. The dependence of $\bm{x}=\dot{\mathbf{r}}-\dot{\mathbf{r}}'$ in the interlayer exchange coupling $\mathcal{J}^{\perp}(\bm{x})$ between spins at in-plane positions $\dot{\mathbf{r}}$ and $\dot{\mathbf{r}}'$. (a) The exchange coupling between AA sublattices in P stacking [see Fig. \ref{fig:Lattice}(c)]. The left panel shows the separate contributions from SE1 and SE2 processes. The right panel is the total exchange coupling. (b) The exchange coupling  between AA sublattices in AP stacking.}
    \label{fig:JInter}
\end{figure}

\begin{figure}
\centering
    \includegraphics[width=3.35in]{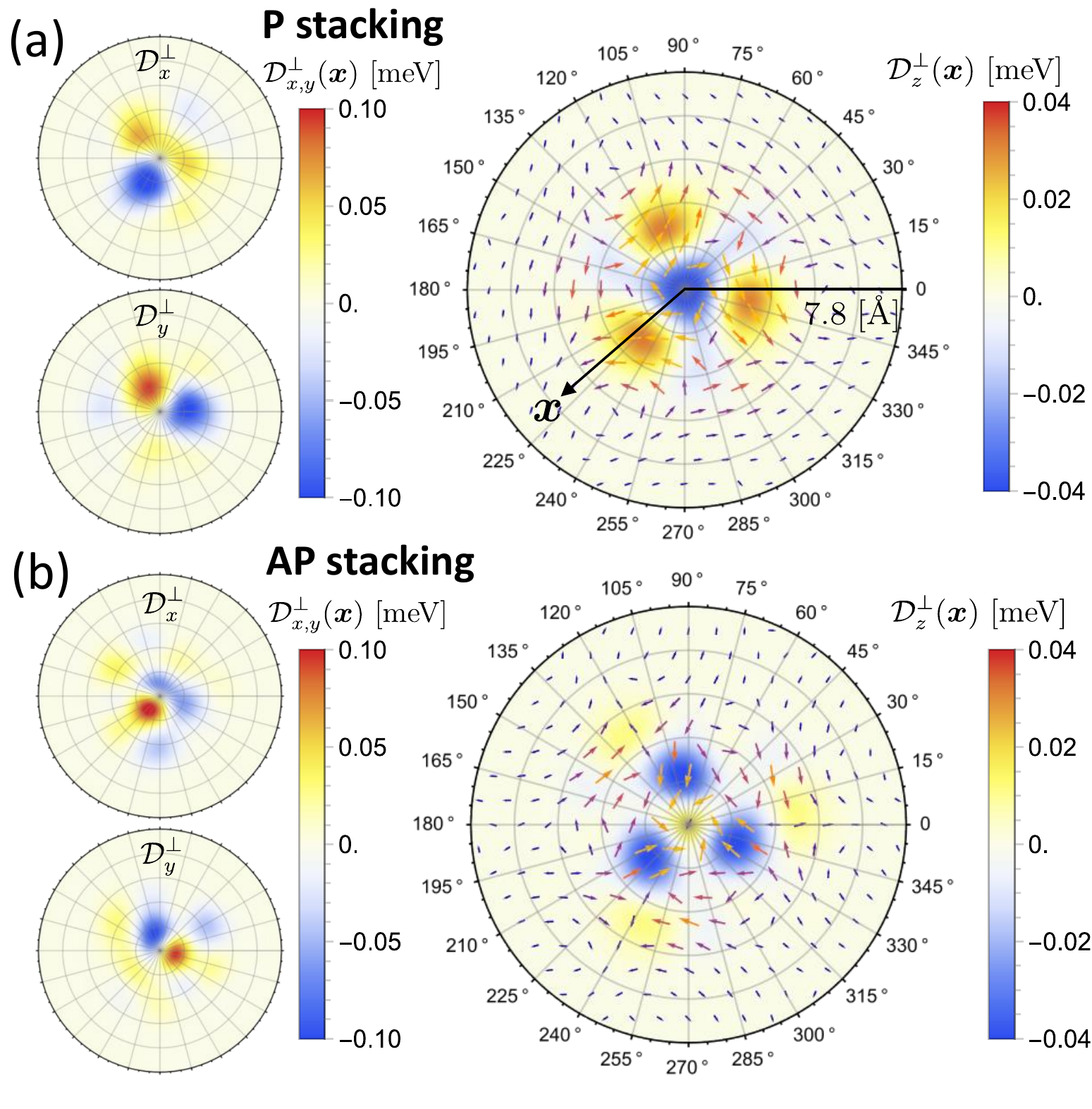}
    \caption{\textbf{Interlayer DM interaction}.(a) Interlayer DM interaction between AA sublattices for P stacking. The left panel is the coupling strength for $\mathcal{D}^{\perp}_x$ and $\mathcal{D}^{\perp}_y$ components. In the right panel, the vector field shows the in-plane component of the DM interaction. The out-of-plane component is illustrated by the color plot. (b) Interlayer DM interaction between AA sublattices in AP stacking. }
    \label{fig:DInter}
\end{figure}

To investigate our model, we analyze the isotropic exchange $\mathcal{J}^{\perp}_{\dot{\mathbf{r}}\dot{\mathbf{r}}'}$ with arbitrary in-plane displacement $\bm{x}=\dot{\mathbf{r}}-\dot{\mathbf{r}}'$ (with $L=6.8$ \AA). In Fig. \ref{fig:JInter}, we plot two different types of superexchange, SE1 and SE2 (left panel). They both contribute to $J^{\perp}_{\dot{\mathbf{r}}\dot{\mathbf{r}}'}$. In the superexchange hopping processes, SE1 creates only \emph{one} virtual $p$ hole ($|\dot{n}n\rangle$) while SE2 creates \emph{two} virtual $p$ holes ($|\dot{m}m,\dot{n}n\rangle_{\xi}$). In contrast to the previous studies, we go beyond the SE1 process and we find that the interlayer AFM exchange is mostly mediated by a virtual singlet hole pair through the $e_g$-$e_g$ hopping process in SE2\cite{Song:PRB106(2022)}. As we can see, SE1 mediates mostly FM exchange which cannot explain the emergence of AFM exchange. Combining SE1 and SE2 gives the total interlayer exchange whose result agrees with the comprehensive DFT calculation in Refs. \cite{Zheng:AdvFuncMat33(2022)} and \cite{Yang:NatCompSci3(2023)}. Similarly, we calculate the interlayer DM interaction in Fig.\ref{fig:DInter}. In P stacking [Fig. \ref{fig:Lattice}(c)], at $\bm{x}=\bm{0}$, we find that $\bm{\mathcal{D}}^{\perp}(0)$ has zero in-plane components due to three-fold rotational symmetry\cite{Moriya:PR120(1960)}. However, in AP stacking, we find that $\bm{\mathcal{D}}^{\perp}(0)$\cite{Moriya:PR120(1960)} vanishes since it has an additional mirror symmetry between the top and bottom layers. Similar to $\mathcal{J}^{\perp}_{\dot{\mathbf{r}}\dot{\mathbf{r}}'}$, we also find that SE1 and SE2 have comparable contributions for $\bm{\mathcal{D}}^\perp_{\dot{\mathbf{r}}\dot{\mathbf{r}}'}$. Furthermore, our theory can also derive the analytical form for the intralayer next-NN DM interaction\footnote{The analytical expression is similar to interlayer DM interaction by replacing $T_{12}^{\ell_1\ell_2}\to v_{12}$. Using $\lambda=0.6$ eV \cite{Lado:2DMat4(2017)}, we estimate the interactions between AA sublattices through the $h_2$-$h_5$ hopping $\bm{\mathcal{D}}'_{AA}=( -0.05, 0.01, -0.09)$, and between BB sublattices $h_1$-$h_6$ hopping $\bm{\mathcal{D}}'_{BB}=( 0.05, -0.01, 0.09)$ in meV [see Fig. \ref{fig:Lattice}(a)]. The result is consistent with Refs.\cite{Xu:PRB101(2020),Kvashnin:PRB102(2020)}. The DM vectors are pointing along the $X$-$X$ bond direction.} $\bm{\mathcal{D}}'$ and the NN symmetric exchange tensor\footnote{The analytical result can be found in SM\cite{SM}.} $\mathcal{K}^{ij}$. These higher-order intralayer exchanges are relevant for topological magnetic phenomena\cite{Chen:PRX8(2018),Chen:PRX11(2021),Mook:PRX11(2021),Cai:PRB104(2021),Kvashnin:PRB102(2020),JaeschkeUbiergo:PRB103(2021),McClarty:AnnRewCondMat13(2022)} and Kitaev physics\cite{Xu:npjCompMat4(2018),Xu:PRL124(2020),Xu:PRB101(2020),Lee:PRL124(2020),Yadav:arXiv2022}.

\section{Magnetic moir\'e bilayer}
\begin{figure}
    \centering
    \includegraphics[width=3.3in]{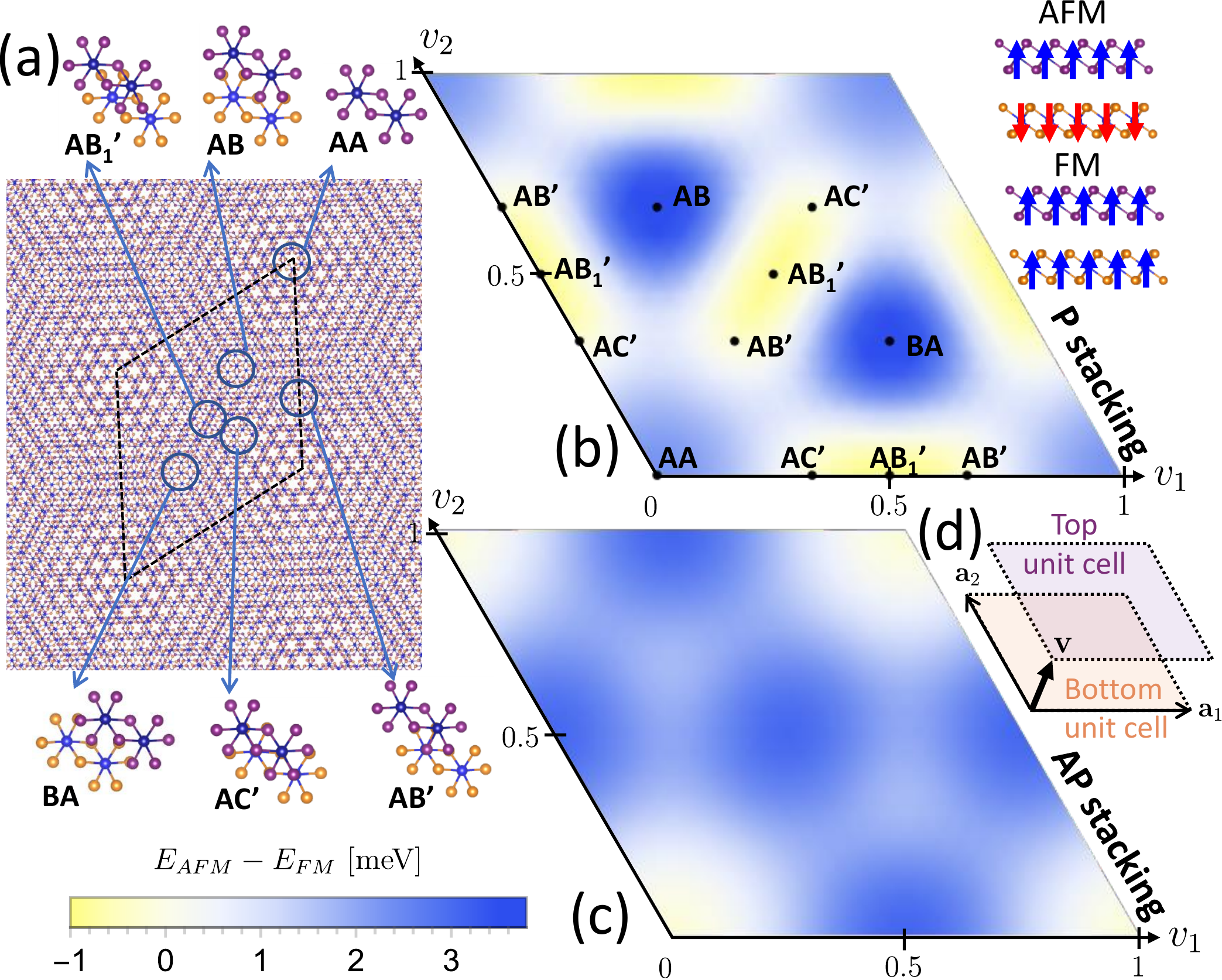}
    \caption{\textbf{Moir\'e field}. (a) The mapping between the local stacking order in a moir\'e lattice and the bilayer stacking with various sliding vectors $\mathbf{v}$. (b) The moir\'e field for P stacking is obtained by using Eqs. \eqref{eqn:Hs} and \eqref{eqn:J_inter}, and setting $\mathcal{K}^{jj'}=0$. (c) The moir\'e field for AP stacking. (d) The unit cells in different layers with the sliding vector $\mathbf{v}=v_1 \mathbf{a}_1+v_2\mathbf{a}_2$.}
    \label{fig:stackingM}
\end{figure}
As demonstrated in previous work\cite{Xiao:PRRes3(2021),Akram:NanoLett21(2021),Fumega:2DM10(2023),Yang:NatCompSci3(2023)}, $\mathcal{J}^{\perp}_{\dot{\mathbf{r}}\dot{\mathbf{r}}'}$ can also be deduced by mapping the local Cr-Cr stacking order in a moir\'e cell to the corresponding sliding bilayer [Fig. \ref{fig:stackingM}(a)]. In this ``sliding-mapping" approach\cite{Yang:NatCompSci3(2023)}, the DFT calculation is performed in the sliding bilayer instead of the twisted bilayer with a large moir\'e cell. To compare our model with this approach, we calculate $E_{AFM}-E_{FM}=\frac{2}{N}\sum_{\dot{\mathbf{r}}\dot{\mathbf{r}}'}(\mathcal{J}^{\perp}_{\dot{\mathbf{r}}\dot{\mathbf{r}}'}+\mathcal{J}^{\perp}_{\dot{\mathbf{r}}'\dot{\mathbf{r}}})(\tfrac{3}{2})^2$ for a bilayer  with $N$ unit cells where $E_{AFM}$ and $E_{FM}$ are the total energies per unit cell in layered-AFM and FM states. The result is plotted as a ``moir\'e field" \cite{Akram:NanoLett21(2021)} with a sliding vector $\mathbf{v}$ in Figs. \ref{fig:stackingM}(b) and \ref{fig:stackingM}(c). Even though our model is constructed based on five DFT data points on the AB stacking (Fig. \ref{fig:KS}), our result in Fig. \ref{fig:stackingM} agrees with the comprehensive DFT studies \cite{Sivadas:ACSnano18(2018),Gibertini:JPhysDapp54(2020),Xiao:PRRes3(2021),Akram:NanoLett21(2021)}. We note that the AFM domain has a slight mismatch with a smaller energy gain as compared to the other DFT studies. This discrepancy may be attributed to the lack of variation of $\bm{p}^\alpha_{\mathbf{r}}$ in response to the change of local crystal field in different stacking structures. This is especially evident for the AP stacking moir\'e field [Fig. \ref{fig:stackingM}(c)], as we use Eq. \eqref{eqn:Tinter} deduced from AB stacking (Fig. \ref{fig:KS}) in the calculation. In superexchange theory, the interlayer exchange is expected to strongly depend on the details of $\bm{p}^\alpha_{\mathbf{r}}$, since it governs the overlapping between $p$ orbitals\cite{Goodenough:JPhysChem6(1958),Kanamori:JPhysChem10(1959)}.

\section{Conclusion} We have built a microscopic spin Hamiltonian for a moir\'e bilayer by using the superexchange theory. Although the construction is done by a relatively small DFT simulation, our analytical model yields a similar result as compared to the rigorous DFT studies. Furthermore, we identify the microscopic origin of the interlayer AFM exchange. This exchange is mediated by the singlet hole pairs in the $X$'s ion\cite{Song:PRB106(2022)}. These low-energy SE2 processes play a vital role in interlayer exchanges.

In superexchange, the interlayer exchange coupling is highly sensitive to the interlayer $X$-$X$ hopping geometry, which can lead to different $\bm{x}$-dependent behavior in twisting (see P and AP stackings in Figs. \ref{fig:JInter} and \ref{fig:DInter}). In the sliding-mapping approach\cite{Sivadas:ACSnano18(2018),Gibertini:JPhysDapp54(2020),Xiao:PRRes3(2021),Akram:NanoLett21(2021)},  the moir\'e fields in P and AP stackings\cite{Gibertini:JPhysDapp54(2020)} [Fig. \ref{fig:stackingM}(b) and \ref{fig:stackingM}(c)] will generate two different $\mathcal{J}^{\perp}(\bm{x})$. Also, P and AP stackings do not transform into each other by sliding since this does not introduce a relative rotation between $X$ planes\footnote{Specifically, in Figs. \ref{fig:Lattice}(c) and \ref{fig:Lattice}(d), sliding does not change the relative orientation between the red dashed triangles in the upper and lower layers. As we see in Figs. \ref{fig:JInter} and \ref{fig:DInter}, this can lead to a qualitative change in the interlayer exchange.}. Hence, the sliding method may not realize all the relevant interlayer exchange coupling in a twisted bilayer. Therefore, we argue that our microscopic approach can complement the incomplete part of the sliding method.

In our theory, the KS parametrization in Eq. \eqref{eqn:Tinter} may not be sufficient since it does not take into account the variation of the $p$-orbital quantization axes in different crystal-field environments. To improve our model by including such effects, it requires more DFT simulations with different stacking configurations. This is particularly important for evaluating anisotropic exchanges such as interlayer DM interactions. Another limitation of our approach is that the Coulomb and Hund's interaction in the $p$ orbital cannot be obtained from the DFT Kohn-Sham spectrum since this information is contained in the two-particle spectrum. This may require further effort in evaluating the two-particle spectrum from first-principles studies.

In conclusion, our work provides microscopic insights into the problem of interlayer magnetic exchange interactions. Unlike conventional bulk materials, the $p$-orbital wavefunctions do not have symmetry constraints in the out-of-plane direction. Owing to the absence of this crystalline symmetry, this unique property in 2D materials leads to a fascinatingly rich stacking-dependent magnetism with remarkable tunability.

\begin{acknowledgements}
The author would like to thank Marco Berritta and Stefano Scali for stimulating discussions. The author also thanks Oleksandr Kyriienko and Vladimir Fal'ko for supporting this work through scientific guidance. The author acknowledges the financial support from U.K. EPSRC New Investigator Award under Agreement No. EP/V00171X/1.
\end{acknowledgements}

\bibliography{CrX3}

\clearpage

\setcounter{table}{0}
\renewcommand{\thesection}{ \arabic\section}

\appendix

\begin{widetext}

\section*{Supplemental Material}

\section{DFT computational details}

The electronic structure of $\mathrm{CrI}_3$ is calculated by using Quantum ESPRESSO. In the calculation, we use the Perdew-Burke-Ernzerhof (PBE) pseudopotentials from the standard solid-state pseudopotential efficiency library. The in-plane monolayer lattice parameter is adopted from the experimental value\cite{McGuire:ACScm27(2015)} 6.867 \AA. To eliminate interaction between supercells, we introduce more than 20 \AA\ vacuum between the supercell images in the out-of-plane direction. To obtain the optimal lattice structure, we perform a fixed unit-cell volume relax calculation on the monolayer lattice (ferromagnetic) until the atomic residual force is less than $3\times10^{-4}$ eV/\AA. The numerical integration over the Brillouin zone is done by sampling over $8\times8\times1$ $\Gamma$-centered Monkhorst-Pack grid. The energy cutoff for wavefunction and density is 50 Ry and 450 Ry. After the relax calculation, we stack two monolayers and compute the electronic structure of AB stacking bilayer (no relaxation) for various interlayer distances between Cr-planes (L= 6.3, 6.8, 7.3, 7.8, 8.3\AA) at the ferromagnetic ground state. In the bilayer calculation, we account for the van der Waal interactions by using the optB86 functional. 

To estimate the TB constants, we construct the maximally localized Wannier functions (MLWF) by using WANNIER90. In the calculation, the $p$ and $d$ Wannier orbitals are projected by using the default angular momentum quantization axes in the package. In this coordinate system, the orbital is quantized by the out-of-plane axis $z$ and $x$, $y$ are the in-plane axes. We note that, in this orbital basis, although the Wannier function is maximally localized, the corresponding \textit{ab initio} Hamiltonian does not have a diagonalized (in orbital space) on-site term. In order to extract the TB constants for $\mathcal{H}_0$ and $\mathcal{H}'$, we perform the change of orbital basis to diagonalize the on-site term of the \textit{ab initio} Hamiltonian and give in the MLWF (still remain well localized). After matching the basis, we can make direct identification of the TB constants ($t_{n\dot{n}}$, $u_{\dot{n}\dot{m}}$, and $v_{nm}$), the $p$ orbital quantization axes ($\bm{p}^{\alpha}_{\mathbf{r}}$), and the quasiparticles energies ($\omega_{\dot{n}}$, $\bar{\omega}_{\dot{n}}$ and $\mathcal{E}_{\dot{n}n}$) with the \textit{ab initio} Hamiltonian\cite{Song:PRB106(2022)}. The intralayer $d$-$p$ ($t_{n\dot{n}}$) and $d$-$d$ ($u_{\dot{n}\dot{m}}$) hopping constants for the AB bilayer are similar to the monolayer TB constants in Ref\cite{Song:PRB106(2022)}. Here, we only summarize some of the intralayer nearest-neighbor $p$-$p$ hopping constants in Table\ref{tbl:TBpp}. They are needed for intralayer next-NN DM interaction and NN symmetric exchange tensor calculations.

\section{Perturbative expansion and the spin Hamiltonian}

To evaluate the partition function, we calculate the expectation value perturbatively in the following by using cumulant expansion\cite{Shankar:RMP66(1994)}
\begin{align}\label{eqn:connect}
\langle\tilde{\Psi}|
&\mathcal{T}e^{-\int_0^\beta d\tau \mathcal{H}_t(\tau)}|\tilde{\Psi}\rangle\!= \!\exp \left(\sum_{n=0}^{\infty} \! \frac{C_n}{n!}\right),
\end{align}
where $C_n$ is the \textit{connected} $n$-correlation functions\cite{Song:PRB106(2022)} ($n$-order cumulants) with
\begin{align}\label{eqn:Cn}
C_2&=\mu_2,\quad C_4=\mu_4-3 \mu_2^2, \quad C_5= \mu_5,\quad 
C_6=\mu_6-15 \mu_4\mu_2+30\mu_2^3,
\end{align}
where the $n$-th moment is defined as
\begin{equation}
\mu_n=\langle\tilde{\Psi}| \mathcal{T}[-\int^\beta_0 d \tau \mathcal{H}'(\tau)]^n|\tilde{\Psi}\rangle.
\end{equation} 
In this letter, we only consider the correlation function $C_n$ up to $n=7$ which gives corrections to the ground state energy. These corrections terms are spin-dependent which leads to the spin Hamiltonian [Eq. \eqref{eqn:connect}] in the low-temperature limit ($\beta\to\infty$).
\begin{equation}
H_s=-\frac{1}{\beta}\sum_{n=2}^{7}\frac{1}{n!}C_n.
\end{equation}
The relation of $C_n$ with the exchange couplings in $H_s$ are summarized as follows
\begin{subequations}
\begin{align}\label{eqn:Cn-J}
\frac{1}{2!}C_2+\frac{1}{4!}C_4=&\beta\sum_{\dot{\mathbf{r}}\dot{\mathbf{r}}',\ell}\mathcal{J}\mathbf{S}^\ell_{\dot{\mathbf{r}}}\cdot\mathbf{S}^\ell_{\dot{\mathbf{r}}'},\\
\frac{1}{5!}C_5=&\beta\sum_{\dot{\mathbf{r}}\dot{\mathbf{r}}',\ell}\bm{\mathcal{D}}\cdot\mathbf{S}^\ell_{\dot{\mathbf{r}}}\times\mathbf{S}^\ell_{\dot{\mathbf{r}}'},\\
\frac{1}{6!}C_6=\frac{1}{6!}[C^{\mathcal{K}}_{6}+C^{\perp}_{6}]
=&\beta\sum_{\dot{\mathbf{r}}\dot{\mathbf{r}}',\ell}\Big[\sum_{jj'}S^\ell_{\dot{\mathbf{r}}j}\mathcal{K}_{jj'}S^\ell_{\dot{\mathbf{r}}'j'}+\sum_{\ell'\neq\ell}\mathcal{J}^{\perp}_{\dot{\mathbf{r}}\dot{\mathbf{r}}'}\mathbf{S}^\ell_{\dot{\mathbf{r}}}\cdot\mathbf{S}^{\ell'}_{\dot{\mathbf{r}}'}\Big],\label{eqn:C6}\\
\frac{1}{7!}C_7=\frac{1}{7!}[C^{\bm{\mathcal{D}}'}_7+C^{\bm{\mathcal{D}}^{\perp}}_7]
=&\beta\sum_{\dot{\mathbf{r}}\dot{\mathbf{r}}',\ell}\Big[\bm{\mathcal{D}}'\cdot\mathbf{S}^\ell_{\dot{\mathbf{r}}}\times\mathbf{S}^\ell_{\dot{\mathbf{r}}'}+\sum_{\ell'\neq\ell}\bm{\mathcal{D}}^\perp_{\dot{\mathbf{r}}\dot{\mathbf{r}}'}\cdot\mathbf{S}^\ell_{\dot{\mathbf{r}}}\times\mathbf{S}^{\ell'}_{\dot{\mathbf{r}}'}\Big].\label{eqn:C7}
\end{align}
\end{subequations}
To calculate $C_n$, we may not need to evaluate all the $n$-th moment $\mu_n$ in full since there are many cancellations in Eq. \eqref{eqn:Cn}. For each $\mu_n$, it contains two different nonzero contributions to $C_n$ depending on the hopping processes. One set of such processes corresponds to a \emph{closed} and \emph{connected} hopping path on the lattice, while the other set corresponds to several \emph{closed} hopping paths but \emph{disconnected}. Here, \emph{connected} means the hopping path is \emph{continuous}. Also, \emph{closed} means the starting point and ending point in the hopping path are the same.\cite{Song:PRB106(2022)}
The correlation function $C_n$ can be obtained by only considering the \emph{connected} and \emph{closed} path processes in $\mu_n$, since the cancellation between the \emph{disconnected} and \emph{closed} path processes in $C_n$ is guaranteed by the link-cluster theorem. This greatly simplifies the $C_n$ calculation. In the next section, we will discuss how to obtain these exchange coupling in Eqs. \eqref{eqn:Cn-J}-\eqref{eqn:C7} from the $C_n$ calculations.

\begin{table}
\caption{\label{tbl:DFTpara}The quasiparticle excitations extracted from DFT calculation ($E_F=0$ is the Fermi energy). Unless stated, the unit is in eV. In this table, the spin up (down) state is parallel (antiparallel) to the FM ground state. The high-energy antiparallel spin $d$-electron states are ignored in the main text. The scissor corrected results are given in the parenthesis.}
\begin{ruledtabular}
\begin{tabular}{lcccccc}
Excitations: & Spin up & & & Spin down &(high-energy) &\\
\hline
$p$-hole : & $\alpha=p_{\tilde{z}},$ & $p_{\tilde{y}},$ & $p_{\tilde{x}}$ & $\alpha=p_{\tilde{z}},$ & $p_{\tilde{y}},$ & $p_{\tilde{x}}$
\\
$\bar{\nu}_{\mathbf{r}\alpha}$ & $-0.221$, & $-0.828$, & $-0.985$ & $-0.256$, & $-0.785$, & $-0.928$
\\
\hline
$d$-hole ($a_{1g}$, $e^{\pi}_g$): & $\dot{\alpha}=1,$ & 2, & 3 & $\dot{\alpha}=1,$ & 2, & 3
\\
$\bar{\omega}_{\dot{\mathbf{r}}\dot{\alpha}}$&$-0.024$, &$0$,&$0$  & $2.969$  (4.169) & $2.956$ (4.156) & $2.956$ (4.156)
\\
$d$-electron ($e^{\sigma}_g$): & $\dot{\alpha}=4,$ & 5, & & $\dot{\alpha}=4,$ & 5, &\\
$\omega_{\dot{\mathbf{r}}\dot{\alpha}}$ &$0.823$, &$0.823$& &$2.922$, &$2.922$&\\
(scissor corrected) &$(2.023)$, &$ (2.023)$& &$(4.122)$, &$ (4.122)$&\\
\end{tabular}
\end{ruledtabular}
\end{table}

\begin{table*}
\caption{\label{tbl:TBpp}The \textit{ab initio} TB constants for the nearest-neigbhor $X$-$X$ hopping: $v_{\dot{\mathbf{r}}\dot{\alpha},\mathbf{r}\alpha}$ [eV]. These TB constants are obtained after the change of orbital basis that gives a diagonalized on-site Hamiltonian. We only present the $h_3\leftrightarrow h_6$ hopping and $h_4\leftrightarrow h_1$ hopping (see Fig.1a in the main text) to show the typical $p$-$p$ hopping strength.}
\begin{ruledtabular}
\begin{tabular}{c|ccccccc}
 Hopping &          & $h_3\to h_6$&         &&            & $h_4\to h_1$&          \\
\hline
  $v_{\mathbf{r}\alpha,\mathbf{r}'\alpha'}$ &$ p_{\tilde{z}}$&$p_{\tilde{x}}$&$p_{\tilde{y}}$&& $ p_{\tilde{z}}$&$p_{\tilde{x}}$&$p_{\tilde{y}}$\\
\hline
$p_{\tilde{z}}$ &-0.2300 & 0.2258 & 0.1937& &-0.2301 & -0.2195 & 0.1921 \\
$p_{\tilde{x}}$&-0.2192 & 0.1701 & 0.1661  & &0.2254 & 0.1701 & -0.1700  \\
$p_{\tilde{y}}$& 0.1920 & -0.1703 &-0.2586   && 0.1931 & 0.1664 & -0.2586  \\
\end{tabular}
\end{ruledtabular}
\end{table*}

\section{Elementary excitations in Mott insulator}

To evaluate $C_n$, we first identify all the relevant excitations and listed  them as follows.

\textbf{Quasiparticle excitations:}
\begin{equation}
\begin{array}{lll}
\text{one $d$-electron:} & |\dot{n}\rangle= \chi^{\ell+}_{\dot{\mathbf{r}}_n\sigma}d^{\ell\dagger}_{\dot{n}\sigma}|\tilde{\Psi}\rangle & \text{with energy $\omega_{\dot{n}}$, $\mathcal{H}_0|\dot{n}\rangle=\omega_{\dot{n}}|\dot{n}\rangle$}.\\
\text{one $d$-hole:} & \bar{\chi}^{\ell+}_{\dot{\mathbf{r}}_n\sigma}d^{\ell}_{\dot{n}\sigma}|\tilde{\Psi}\rangle & \text{with energy $\bar{\omega}_{\dot{n}}$, $\mathcal{H}_0\bar{\chi}^\sigma_{\dot{\mathbf{r}}_n+}d^{\ell}_{\dot{n}\sigma}|\tilde{\Psi}\rangle=\bar{\omega}_{\dot{n}}\bar{\chi}^\sigma_{\dot{\mathbf{r}}_n+}d_{\dot{n}\sigma}|\tilde{\Psi}\rangle$}.\\
\text{one $p$-electron:} & \chi^{\ell+}_{\dot{\mathbf{r}}_n\sigma'} p^{\ell\dagger}_{n \sigma'}|\tilde{\Psi}\rangle=0 & \text{$p$ orbitals are fully occupied.}\\
\text{one $p$-hole:} & |n\rangle=\bar{\chi}^{\ell+}_{\dot{\mathbf{r}}_n\sigma'} p^{\ell}_{n \sigma'}|\tilde{\Psi}\rangle & \text{with energy $\bar{\nu}_{n}$, $\mathcal{H}_0|n\rangle=\bar{\nu}_{n}|n\rangle$}.
\end{array}
\end{equation}
In the above, the quasiparticle energies can be obtained from Table \ref{tbl:DFTpara}. In this letter, we only consider the excitations with spin parallel to the FM ground state (spin-up states). The high-energy excitations with spin antiparallel to the FM ground states (spin-down states) are omitted. These antiparallel spin excitations in the $d$ orbitals are complicated due to Hund's interaction\cite{Song:PRB106(2022)}. One can take into account such excitations. However, their contributions are strongly suppressed by a large energy gap.

In virtual hopping processes, the elementary excitations are electron-hole pairs since the TB hopping Hamiltonian preserves the total number of particles. Here, we list these relevant excitations as follows

\textbf{Electron-hole pairs excitations:}
\begin{equation}
\begin{array}{lll}
\text{one electron-hole pair:} & |\dot{n}n\rangle= \bar{\chi}^{\ell+}_{\dot{\mathbf{r}}_n\sigma'} p^{\ell}_{n \sigma'}|\dot{n}\rangle& \text{with energy $\mathcal{E}_{\dot{n}n}=\omega_{\dot{n}}-\bar{\nu}_n$}.\\
\text{two electron-hole pairs:} &|\dot{m}m,\dot{n}n\rangle_{\xi}=\bar{\chi}^{\ell+}_{\dot{\mathbf{r}}_m\sigma'}\bar{\chi}^{+}_{\dot{\mathbf{r}}_n\sigma''} (p^\ell_{m \sigma'}p^\ell_{n \sigma''}\!+\!\xi p^\ell_{m \sigma''}p^\ell_{n \sigma'})|\dot{m}\dot{n}\rangle & \text{with energy $\mathcal{E}_{\dot{n}n}\!+\!\mathcal{E}_{\dot{m}m}\!+\!\Theta_{mn}^\xi$}.
\end{array}
\end{equation}
where $\xi=\pm1$ corresponds to the singlet/triplet state in the $p$-holes. Because of the interaction between two $p$ holes, this leads to the singlet-triplet splitting with energy $\Theta^\xi_{mn}=\frac{1}{2}[U^{\alpha_m \alpha_n}_{\mathbf{r}_m}-(2 \xi+1)(1-\delta_{\alpha_m \alpha_n})J^{\alpha_m \alpha_n}_{\mathbf{r}_m}]\delta_{\mathbf{r}_m \mathbf{r}_n}$.

One may express these quasiparticle energies explicitly in terms of the model parameters\cite{Song:PRB106(2022)} in $\mathcal{H}_0$. However, these are not necessary, since these quasielectron and quasihole energies can be directly obtained from the \textit{ab inito} TB Hamiltonian (in Table\ref{tbl:DFTpara}). Once these low-energy excitations of the Mott's states are identified. The rest of the calculation is merely a straightforward algebra in perturbation expansion.
 
\section{Imaginary-time evolution of the hopping processes.}
With the excitations energies in the previous section, we calculate the imaginary-time evolving states which will be needed for evaluating $C_n$. The relevant excitations with $n$-time processes are listed below. In the rest of this supplemental material, we implicitly assumed the summation of all orbital indices ($\dot{\alpha}$, $\alpha$), position indices ($\dot{\mathbf{r}}$, $\mathbf{r}$), and spin index ($\sigma$), unless otherwise stated. Also, we let
\begin{align}
\mathcal{H}_t=&\sum_{\ell \dot{1}1 }(t_{1\dot{1}}p^{\ell\dagger}_{1 \sigma}d^{\ell}_{\dot{1} \sigma}+t_{\dot{1}1}d^{\ell\dagger}_{\dot{1} \sigma}p^{\ell}_{1 \sigma}),\\
\mathcal{V}_{\lambda}=&\sum_{\ell, 12}\bm{\Lambda}_{12}\cdot\bm{\tau}_{\sigma\sigma'}p^{\ell\dagger}_{1 \sigma}p^{\ell}_{2\sigma'},\\
\mathcal{H}_{\perp}=&\sum_{\ell\ell', 12}\!T^{\ell\ell'}_{12}\delta_{\sigma \sigma'}p^{\ell\dagger}_{1 \sigma}p^{\ell'}_{2\sigma'}.
\end{align}

\textbf{Two-time processes:}
\begin{align}
	\mathcal{H}_t(\tau_2)\mathcal{H}^\dagger_t(\tau_1)|\tilde{\Psi}\rangle=&\bar{\chi}^{\ell}_{ \dot{\mathbf{r}}_1\sigma}\chi^{\ell}_{ \dot{\mathbf{r}}_2\sigma}P_{ \dot{\alpha}_2}t_{\dot{1} 1}(\tau_1)  \bar{t}_{ 1 \dot{2} }(\tau_2)
\mathsf{d}^\ell_{\dot{2} }
\mathsf{d}^{\ell\dagger}_{\dot{1}} 
|\tilde{\Psi}\rangle,\quad\quad\text{(with $\mathsf{d}_{\dot{n}}^\ell=\bar{\chi}^{\ell+}_{\dot{\mathbf{r}}_n \sigma}d_{\dot{n} \sigma}^\ell$)}\label{eqn:HH+}\\
	\mathcal{H}^\dagger_t(\tau_2)\mathcal{H}^\dagger_t(\tau_1)|\tilde{\Psi}\rangle=&\bar{X}^{\sigma_1 \sigma_2}_{\ell\dot{\mathbf{r}}_1,\ell\dot{\mathbf{r}}_2,\xi}
   t_{\dot{1} 1} (\tau_1)t_{\dot{2} 2} (\tau_2)\mathrm{e}^{\tau_2 \Theta^{\xi }_{12}}p^\ell_{ 2\sigma_2}p^\ell_{ 1\sigma_1}
\mathsf{d}^{\ell\dagger}_{\dot{2}} \mathsf{d}^{ \ell\dagger}_{\dot{1}} 
|\tilde{\Psi}\rangle,\label{eqn:HdHd}
	\\
	\mathcal{V}_\lambda(\tau_2)\mathcal{H}^\dagger_t(\tau_1)|\tilde{\Psi}\rangle=&
 \bar{\slashed{\chi}}^{\ell,j}_{ \dot{\mathbf{r}}_1\sigma_2}\Lambda^j_{12}(\tau_2)t_{\dot{1} 1} (\tau_1) p^\ell_{ 2 \sigma_2}
\mathsf{d}^{\ell\dagger}_{\dot{1} }
|\tilde{\Psi}\rangle,
\\
\mathcal{H}_\perp(\tau_2)\mathcal{H}^\dagger_t(\tau_1)|\tilde{\Psi}\rangle=&\bar{\chi}^{\ell}_{ \dot{\mathbf{r}}_1\sigma_1}
t_{\dot{1}1}(\tau_1) T^{\ell_1\ell_2}_{12}(\tau_2)p^{\ell_2}_{2 \sigma_1}\mathsf{d}^{\ell_1\dagger}_{\dot{1} }|\tilde{\Psi}\rangle,
\end{align}
where $P_{\dot{\alpha}}=1$ for $\dot{\alpha}=1,2,3$ and $P_{\dot{\alpha}}=0$ for $\dot{\alpha}=4,5$. In the above, we only keep the low-energy excitations that are described in the previous section. Furthermore, we dropped the superscript `+' in the spin wavefunction $\chi^{+}_{\dot{\mathbf{r}}\sigma}$ for simplicity. 
Also, we let $t_{ \dot{1} 1}(\tau_1)=\mathrm{e}^{\tau_1 \omega_{\dot{1}}}t_{  \dot{1}1 } \mathrm{e}^{-\tau_1 \bar{\nu}_{ 1}}$, $\bar{t}_{ 1\dot{1} }(\tau_1)=\mathrm{e}^{\tau_1 \bar{\nu}_{ 1}}t_{ 1 \dot{1} } \mathrm{e}^{-\tau_1 \bar{\omega}_{\dot{1}}}$, and $\Lambda^j_{12}(\tau_2)=\mathrm{e}^{\tau_2 \bar{\nu}_{\mathbf{r}_1\alpha_1}}[\bm{\Lambda}_{12}]_j\mathrm{e}^{-\tau_2 \bar{\nu}_{\mathbf{r}_1\alpha_2}}$. Also, we use the Dirac slash notation $
\bar{\slashed{\chi}}^{\ell, j}_{ \dot{\mathbf{r}}_1\sigma_2}= \bar{\chi}^{\ell,+}_{ \dot{\mathbf{r}}_1\sigma_1}\tau^j_{\sigma_1 \sigma_2}$ to contract the spin-sum, where $\bm{\tau}=(\tau^x,\tau^y,\tau^z)$. The singlet-triplet energy splitting is $\Theta^\xi_{mn}=\frac{1}{2}[U^{\alpha_m \alpha_n}_{\mathbf{r}_m}-(2 \xi+1)(1-\delta_{\alpha_m \alpha_n})J^{\alpha_m \alpha_n}_{\mathbf{r}_m}]\delta_{\mathbf{r}_m \mathbf{r}_n}$.
In Eq. \eqref{eqn:HdHd}, we let the spin-singlet and spin-triplet pair in the $d$ orbital as
\begin{equation*}
\bar{X}^{\sigma_1 \sigma_2}_{\ell\dot{\mathbf{r}}_1,\ell'\dot{\mathbf{r}}_2,\xi}=
\frac{1}{2}(\bar{\chi}^{\ell+}_{\dot{\mathbf{r}}_1\sigma_1}\bar{\chi}^{\ell'+}_{ \dot{\mathbf{r}}_2\sigma_2}+\xi\bar{\chi}^{\ell+}_{ \dot{\mathbf{r}}_1\sigma_2}\bar{\chi}^{\ell'+}_{ \dot{\mathbf{r}}_2\sigma_1}).
\end{equation*}

We can further calculate the higher-order excitations that involve the spin-flip processes from the spin-orbit coupling. These appear in

\textbf{Three-time processes:}
\begin{align}
\mathcal{V}_\lambda(\tau_3)\mathcal{H}^\dagger_t(\tau_2)\mathcal{H}^\dagger_t(\tau_1)|\tilde{\Psi}\rangle\!
=&
\frac{1}{2}\Xi_{\eta}(\tau_1|_{\dot{1}1},\tau_2|_{\dot{2}2})\mathrm{e}^{(\tau_2-\tau_3)\Theta^{\eta }_{12}}\Lambda_{23}^{j}(\tau_3)\mathrm{e}^{\tau_3\Theta^{\eta'}_{13}}
(\bar{\slashed{X}}^{j,\sigma_3\sigma_1}_{\ell\dot{\mathbf{r}}_1,\ell\dot{\mathbf{r}}_2,\eta}+\eta'\bar{\slashed{X}}^{j,\sigma_1\sigma_3}_{\ell\dot{\mathbf{r}}_1,\ell\dot{\mathbf{r}}_2,\eta})p^{\ell}_{ 3\sigma_3}p^{\ell}_{ 1\sigma_1}
\mathsf{d}^{\ell \dagger}_{\dot{2}}\mathsf{d}^{\ell\dagger}_{\dot{1} }|\tilde{\Psi}\rangle,
\\
\mathcal{H}_{t}(\tau_3)\mathcal{V}_\lambda(\tau_2)\mathcal{H}^\dagger_t(\tau_1)|\tilde{\Psi}\rangle\!
=&-\chi^{\ell}_{ \dot{\mathbf{r}}_3\sigma_2}\tau^{j}_{\sigma_1 \sigma_2}\bar{\chi}^{\ell}_{ \dot{\mathbf{r}}_1\sigma_1}P_{ \dot{\alpha}_3}   t_{\dot{1}  1} (\tau_1)\Lambda_{12}^{j}(\tau_2)\bar{t}_{2\dot{3}}(\tau_3)\mathsf{d}_{\dot{3}}\mathsf{d}^{\dagger}_{\dot{1} } |\tilde{\Psi}\rangle,
\\
\mathcal{H}^\dagger_t(\tau_3)\mathcal{V}_\lambda(\tau_2)\mathcal{H}^\dagger_t(\tau_1)|\tilde{\Psi}\rangle\!
=&\frac{1}{2} (\bar{\slashed{\chi}}^{\ell,j}_{ \dot{\mathbf{r}}_1\sigma_2}\bar{\chi}^{\ell}_{\dot{\mathbf{r}}_2\sigma_3}+\eta\bar{\slashed{\chi}}^{\ell,j}_{ \dot{\mathbf{r}}_1\sigma_3}\bar{\chi}^{\ell}_{\dot{\mathbf{r}}_2\sigma_2})t_{\dot{1}  1} (\tau_1) \Lambda_{12}^{j}(\tau_2)t_{ \dot{2}3}(\tau_3)\mathrm{e}^{\tau_3\Theta^{\eta}_{23}} p_{ 3\sigma_3}p_{ 2\sigma_2}\mathsf{d}^{ \dagger}_{\dot{2}}\mathsf{d}^{ \dagger}_{\dot{1}}|\tilde{\Psi}\rangle,
\end{align}
where we let
\begin{equation}
 \Xi_{\eta}(\tau_1|_{\dot{1}1},\tau_2|_{\dot{2}2})=t_{\dot{1}  1} (\tau_1) t_{\dot{2} 2}(\tau_2)- \eta t_{\dot{1}  1} (\tau_1) t_{\dot{2} 2}(\tau_2).
\end{equation}
Here, we also use the Dirac slashed notation
\begin{equation}
\bar{\slashed{X}}^{\sigma_3\sigma_1,j}_{\ell_1\dot{\mathbf{r}}_1,\ell_2\dot{\mathbf{r}}_2,\eta}=\bar{X}^{\sigma_2\sigma_1}_{\ell_1\dot{\mathbf{r}}_1,\ell_2\dot{\mathbf{r}}_2,\eta}\tau^j_{\sigma_2\sigma_3},
\end{equation}
and the excitations created by interlayer hopping
\begin{align}
\mathcal{H}_{\perp}(\tau_3)\mathcal{H}^{\dagger}_{t}(\tau_2)\mathcal{H}^{\dagger}_{t}(\tau_1)|\tilde{\Psi}\rangle\!=
	& 
\bar{X}^{\sigma_1 \sigma_2}_{\ell_1\dot{\mathbf{r}}_1,\ell_2\dot{\mathbf{r}}_2,\eta}
 \Xi_{-1}(\tau_1|_{\dot{1}1},\tau_2|_{\dot{2}2})
T^{\ell_1\ell_2}_{ 13}(\tau_3)\mathrm{e}^{\tau_3\Theta^{\eta }_{23}}
p^{\ell_2}_{ 3\sigma_1}p^{\ell_2}_{ 2\sigma_2}
\mathsf{d}^{\ell_2\dagger}_{ \dot{2}}\mathsf{d}^{\ell_1\dagger}_{\dot{1}}|\tilde{\Psi}\rangle,\\
	\mathcal{H}_{t}(\tau_3)\mathcal{H}_{\perp}(\tau_2)\mathcal{H}^{\dagger}_{t}(\tau_1)|\tilde{\Psi}\rangle\!=&
 -P_{\dot{\alpha}_2}t_{\dot{1}  1} (\tau_1)T^{\ell_1\ell_2}_{ 1 2}(\tau_2)  \bar{t}_{ 2 \dot{ 2}}(\tau_3)\bar{\chi}^{\ell_1}_{\dot{\mathbf{r}}_1\sigma_1}\chi^{\ell_2}_{\dot{\mathbf{r}}_2\sigma_1}\mathsf{d}^{\ell_2}_{ \dot{2}} \mathsf{d}^{\ell_1\dagger}_{\dot{1}}|\tilde{\Psi}\rangle,\\
	\mathcal{H}^\dagger_t(\tau_3)\mathcal{H}_{\perp}(\tau_2)\mathcal{H}^\dagger_t(\tau_1)|\tilde{\Psi}\rangle\!=&
\eta\bar{X}^{\sigma_1 \sigma_2}_{\ell_1\dot{\mathbf{r}}_1,\ell_2\dot{\mathbf{r}}_2,\eta}
t_{\dot{1} 1}^{ m_1}(\tau_1)T^{\ell_1\ell_2}_{1 2}(\tau_2) t_{ \dot{2} 3}(\tau_3) \mathrm{e}^{\tau_3\Theta^{\eta}_{23}}
p^{\ell_2}_{ 3 \sigma_1}p^{\ell_2}_{ 2 \sigma_2} \mathsf{d}^{\ell_2\dagger}_{ \dot{2}}\mathsf{d}^{\ell_1\dagger}_{ \dot{1}}|\tilde{\Psi}\rangle.\label{eqn:H+H''H+}
\end{align}

Finally, we then proceed to the four-time processes. This is necessary for obtaining the interlayer DM interactions and the intralayer next-NN DM interaction. They are listed as following

\textbf{Four-time processes:}
\begin{align}
	\mathcal{H}_{\perp}(\tau_4)\mathcal{V}_\lambda(\tau_3)\mathcal{H}^\dagger_t(\tau_2)\mathcal{H}^\dagger_t(\tau_1)|\tilde{\Psi}\rangle\!
=&
\Big\{
(\bar{\chi}^{\sigma_1}_{\ell_1\dot{\mathbf{r}}_1\dot{\sigma}_1}\bar{\slashed{\chi}}^{j,\sigma_2}_{\ell_2\dot{\mathbf{r}}_2\dot{\sigma}_2}+\eta\bar{\chi}^{\sigma_2}_{\ell_1\dot{\mathbf{r}}_1\dot{\sigma}_1}\bar{\slashed{\chi}}^{j,\sigma_1}_{\ell_2\dot{\mathbf{r}}_2\dot{\sigma}_2})\Xi_{-1}(\tau_1|_{\dot{1}1},\tau_2|_{\dot{2}3})
 \Lambda^{j}_{32}(\tau_3)\notag\\
&
+\eta(\bar{\chi}^{\sigma_1}_{\ell_2\dot{\mathbf{r}}_2\dot{\sigma}_2}\bar{\slashed{\chi}}^{j,\sigma_2}_{\ell_1\dot{\mathbf{r}}_1\dot{\sigma}_1}+\eta\bar{\chi}^{\sigma_2}_{\ell_2\dot{\mathbf{r}}_2\dot{\sigma}_2}\bar{\slashed{\chi}}^{j,\sigma_1}_{\ell_1\dot{\mathbf{r}}_1\dot{\sigma}_1})\Xi_{-1}(\tau_1|_{\dot{2}2},\tau_2|_{\dot{1}3})
 \Lambda^{j}_{31}(\tau_3)
\Big\}
\notag\\
&T^{\ell_1\ell_2}_{14}(\tau_4)\mathrm{e}^{\tau_4\Theta^\eta_{24}}p^{\ell_2}_{ 4\sigma_1}p^{\ell_2}_{2\sigma_2}
\mathsf{d}^{\ell_2\dagger}_{ \dot{2}}\mathsf{d}^{\ell_1\dagger}_{\dot{1}}|\tilde{\Psi}\rangle,
\\
	\mathcal{V}_{\lambda}(\tau_4)\mathcal{H}^\dagger_t(\tau_3)\mathcal{H}_{\perp}(\tau_2)\mathcal{H}^\dagger_t(\tau_1)|\tilde{\Psi}\rangle
=&
-
\eta(\bar{\slashed{X}}^{j,\sigma_1 \sigma_2}_{\ell_1\dot{\mathbf{r}}_1,\ell_2\dot{\mathbf{r}}_2,\eta}+\eta'\bar{\slashed{X}}^{j,\sigma_2 \sigma_1}_{\ell_1\dot{\mathbf{r}}_1,\ell_2\dot{\mathbf{r}}_2,\eta})
t_{\dot{1} 1}^{ m_1}(\tau_1)[T^{\ell_1\ell_2}_{1 2}(\tau_2) t_{ \dot{2} 3}(\tau_3) -\eta T^{\ell_1\ell_2}_{1 3}(\tau_2) t_{ \dot{2} 2}(\tau_3) ]\notag\\
&
\mathrm{e}^{(\tau_3-\tau_4)\Theta^{\eta}_{23}}\mathrm{e}^{\tau_4\Theta^{\eta'}_{24}}\Lambda_{34}^{j}(\tau_4) p^{\ell_2}_{4 \sigma_1}p^{\ell_2}_{ 2 \sigma_2}
 \mathsf{d}^{\ell_2\dagger}_{ \dot{2}}\mathsf{d}^{\ell_1\dagger}_{ \dot{1}}|\tilde{\Psi}\rangle,\\
	\mathcal{V}_{\lambda}(\tau_4)\mathcal{H}_{\perp}(\tau_3)\mathcal{H}^{\dagger}_{t}(\tau_2)		\mathcal{H}^{\dagger}_{t}(\tau_1)|\tilde{\Psi}\rangle\!
=&
-(\bar{\slashed{X}}^{j,\sigma_1 \sigma_2}_{\ell_1\dot{\mathbf{r}}_1,\ell_2\dot{\mathbf{r}}_2,\eta}+\eta'\bar{\slashed{X}}^{j,\sigma_2 \sigma_1}_{\ell_1\dot{\mathbf{r}}_1,\ell_2\dot{\mathbf{r}}_2,\eta})
\Big\{\Xi_{-1}(\tau_1|_{\dot{1}1},\tau_2|_{\dot{2}2})
T^{\ell_1\ell_2}_{ 13}(\tau_3)\notag\\
&
-\eta
\Xi_{-1}(\tau_1|_{\dot{1}1},\tau_2|_{\dot{2}3})
T^{\ell_1\ell_2}_{ 12}(\tau_3)\Big\}\mathrm{e}^{(\tau_3-\tau_4)\Theta^{\eta }_{23}}\Lambda_{34}^{j}(\tau_4)\mathrm{e}^{\tau_4\Theta^{\eta'}_{24}} p^{\ell_2}_{4 \sigma_1}p^{\ell_2}_{ 2\sigma_2}
\mathsf{d}^{\ell_2\dagger}_{ \dot{2}}\mathsf{d}^{\ell_1\dagger}_{\dot{1}}|\tilde{\Psi}\rangle,
\\
	\mathcal{H}^\dagger_t(\tau_4)\mathcal{V}_{\lambda}(\tau_3)\mathcal{H}_{\perp}(\tau_2)\mathcal{H}^\dagger_t(\tau_1)|\tilde{\Psi}\rangle\!=&
 -
 \bar{\slashed{X}}^{j, \sigma_2\sigma_1}_{\ell_1\dot{\mathbf{r}}_1,\ell_2\dot{\mathbf{r}}_2,\eta}
t_{\dot{1}1}(\tau_1) T^{\ell_1\ell_2}_{13}(\tau_2)\Lambda_{32}^{j}(\tau_3)t_{\dot{2}4}(\tau_4)\mathrm{e}^{\tau_4 \Theta^{\eta}_{24}} p^{\ell_{2}}_{ 4 \sigma_1}p^{\ell_2}_{2 \sigma_2}\mathsf{d}^{\ell_2\dagger}_{\dot{2} }\mathsf{d}^{\ell_1\dagger}_{\dot{1} }|\tilde{\Psi}\rangle,\\
	\mathcal{H}^\dagger_t(\tau_4)\mathcal{H}_{\perp}(\tau_3)\mathcal{V}_{\lambda}(\tau_2)\mathcal{H}^\dagger_t(\tau_1)|\tilde{\Psi}\rangle\!=&
-\bar{\slashed{X}}^{ j,\sigma_2\sigma_1}_{\ell_1\dot{\mathbf{r}}_1,\ell_2\dot{\mathbf{r}}_2,\eta}t_{\dot{1} 1} (\tau_1)\Lambda_{13}^{j}(\tau_2)T^{\ell_{1}\ell_2}_{32}(\tau_3)t_{\dot{2}4}(\tau_4)\mathrm{e}^{\tau_4 \Theta^{\eta}_{24}} p^{\ell_{2}}_{ 4 \sigma_1}p^{\ell_{2}}_{ 2 \sigma_2}\mathsf{d}^{\ell_2\dagger}_{\dot{2} }
\mathsf{d}^{\ell_1\dagger}_{\dot{1} }|\tilde{\Psi}\rangle,\\
	\mathcal{H}_{t}(\tau_4)\mathcal{V}_{\lambda}(\tau_3)\mathcal{H}_{\perp}(\tau_2)\mathcal{H}^{\dagger}_{t}(\tau_1)|\tilde{\Psi}\rangle\!=&
 -
 \bar{\slashed{\chi}}^{j,\sigma_3}_{ \dot{\mathbf{r}}_1\dot{\sigma}_1}\chi^{\sigma_3}_{\mathbf{r}_2+}P_{\dot{\alpha}_2}
t_{\dot{1}1}(\tau_1) T^{\ell_1\ell_2}_{12}(\tau_2)\Lambda_{23}^{j}(\tau_3)t_{3\dot{2}}(\tau_4)\mathsf{d}^{\ell_2}_{\dot{2}} \mathsf{d}^{\ell_1\dagger}_{\dot{1} }|\tilde{\Psi}\rangle,\\
	\mathcal{H}_{t}(\tau_4)\mathcal{H}_{\perp}(\tau_3)\mathcal{V}_{\lambda}(\tau_2)\mathcal{H}^{\dagger}_{t}(\tau_1)|\tilde{\Psi}\rangle\!=&
-\bar{\slashed{\chi}}^{j,\sigma_2}_{ \dot{\mathbf{r}}_1\dot{\sigma}_1}\chi^{\sigma_2}_{\mathbf{r}_2+}P_{\dot{\alpha}_2}t_{\dot{1} 1} (\tau_1)\Lambda_{12}^{j}(\tau_2)T^{\ell_{1}\ell_2}_{23}(\tau_3)t_{3\dot{2}}(\tau_4)\mathsf{d}^{\ell_2}_{\dot{2}} 
\mathsf{d}^{\ell_1\dagger}_{\dot{1} }
|\tilde{\Psi}\rangle.\label{eqn:HHpVH+}
\end{align}
These processes are sufficient for performing the perturbative expansion up to $n=7$.

\section{Superexchange calculations}

In this section, we provide the detail for calculating the higher-order perturbative corrections to Mott's insulating ground states. In the calculation, the low-temperature limit $\beta\to\infty$ is applied to all final results.

\subsection{Tools for $C_n$ calculations} 
First, we summarize the basic tools that are used in evaluating the perturbation expansion.

\textbf{1. Imaginary-time integration.}
To perform the  $\tau$-integration, we get rid of the time-ordering operator ($\mathcal{T}$) and write the integral as 
\begin{equation}
\int_0^\beta d\tau_1 \int_0^\beta \dots d \tau_n\mathcal{T}[\dots]=n!\int_0^\beta \!\!d \tau_n\int^{\tau_{n}}_{0}\!\!d \tau_{n-1}\dots\int^{\tau_{2}}_{0}\!\! d \tau_1[\dots],
\end{equation}
with $0<\tau_1 < \dots < \tau_{n} <\beta $
In the following calculation, we use the notation $\int_\tau d(n\dots 21)=\int_0^\beta \!\!d \tau_n\int^{\tau_{n}}_{0}\!\!d \tau_{n-1}\dots\int^{\tau_{2}}_{0}\!\! d \tau_1$. 

\textbf{2. $\sigma$-spin sum.}
The the spin-sum of the spin wavefunction $\bar{\chi}_{ \dot{\mathbf{r}}\sigma}$ and $ \chi_{ \dot{\mathbf{r}}\sigma}$ are
\begin{align}
\bar{\chi}^+_{\dot{\mathbf{r}}\sigma}\bm{\tau}_{\sigma \sigma'}\chi^+_{\dot{\mathbf{r}}\sigma'}
&=
\mathbf{s}_{ \dot{\mathbf{r}}},
\\
|\bar{\chi}^+_{\dot{\mathbf{r}_2}\sigma'}\chi^+_{\dot{\mathbf{r}_1}\sigma}|^2
&=
\frac{1}{2}(1+\mathbf{s}_{ \dot{\mathbf{r}}_2}\cdot\mathbf{s}_{ \dot{\mathbf{r}}_1}),
\\
\bar{\chi}^+_{\dot{\mathbf{r}_1}\sigma_1}\bm{\tau}_{\sigma_1\sigma_2}\chi^+_{\dot{\mathbf{r}_2}\sigma_2}\bar{\chi}^+_{\dot{\mathbf{r}_2}\sigma}\chi^+_{\dot{\mathbf{r}_1}\sigma}
&=
\frac{1}{2}[-\mathbf{s}_{ \dot{\mathbf{r}}_1}-\mathbf{s}_{ \dot{\mathbf{r}}_2}+i\mathbf{s}_{ \dot{\mathbf{r}}_1}\times\mathbf{s}_{ \dot{\mathbf{r}}_2}],
\\
\bar{\chi}^+_{\dot{\mathbf{r}_1}\sigma_1}\tau^j_{\sigma_1\sigma_2}\chi^+_{\dot{\mathbf{r}_2}\sigma_2}\bar{\chi}^+_{\dot{\mathbf{r}_2}\sigma_2'}\tau^{j'}_{\sigma_2'\sigma_1'}\chi^+_{\dot{\mathbf{r}_1}\sigma_1'}
&=
\frac{1}{2}[i\varepsilon_{j j' j''}(s_{ \dot{\mathbf{r}}_1j''}-s_{ \dot{\mathbf{r}}_2j''})+\delta_{j j'}(1-\mathbf{s}_{ \dot{\mathbf{r}}_2}\cdot\mathbf{s}_{ \dot{\mathbf{r}}_1})+(s_{ \dot{\mathbf{r}}_2j}s_{ \dot{\mathbf{r}}_1j'}+s_{ \dot{\mathbf{r}}_2j'}s_{ \dot{\mathbf{r}}_1j})].\label{eqn:tensor-spin}
\end{align}
where, in the above, the sum of the spin index is implicitly assumed except $\dot{\mathbf{r}}_{1,2}$. $\bm{\tau}=[\tau^x,\tau^y,\tau^z]$ are Pauli's matrices.

\textbf{3. Expectation values.} In the calculation, we need the following expectation values
\begin{align}
\langle\tilde{\Psi}|p^{\ell_4\dagger}_{4\sigma_4}p^{\ell_3\dagger}_{3\sigma_3}&p^\ell_{2\sigma_2}p^\ell_{1\sigma_1}|\tilde{\Psi}\rangle=\delta(14)\delta(23)-\delta(13)\delta(24)
\end{align}
with $\delta(12)=\delta_{\mathbf{r}_1\mathbf{r}_2}\delta_{\alpha_1\alpha_2}\delta_{\sigma_1\sigma_2}\delta_{\ell_1\ell_2}$, and
\begin{align}
\langle\tilde{\Psi}|\mathsf{d}^{\ell_4}_{ \dot{4}}\mathsf{d}^{\ell_3}_{ \dot{3}}\mathsf{d}^{\ell_2\dagger}_{\dot{2}}\mathsf{d}^{\ell_1\dagger}_{\dot{1}}|\tilde{\Psi}\rangle=\delta(\dot{3}\dot{2})\delta(\dot{4}\dot{1})-\delta(\dot{3}\dot{1})\delta(\dot{4}\dot{2})
\end{align}
with $\delta(\dot{1}\dot{2})=\delta_{\dot{\mathbf{r}}_1\dot{\mathbf{r}}_2}\delta_{\dot{\alpha}_1\dot{\alpha}_2}\delta_{\ell_1\ell_2}$.

\textbf{4. $\tau$-dependent conjugate states.} The $\tau$-evolution of the excited states in Eqs. \eqref{eqn:HH+} to \eqref{eqn:HHpVH+} can be transform into the conjugate states as follow
\begin{equation}
\langle\tilde{\Psi}|\mathcal{H}'(\tau_n)\dots\mathcal{H}'(\tau_1)\!=\!\Big[\mathcal{H}'(-\tau_1)\dots\mathcal{H}'(-\tau_n)|\tilde{\Psi}\rangle\Big]^\dagger.
\end{equation}
This is used for calculating the expectation in $C_n$. 

\subsection{Interlayer exchange}
We begin with the discussion of interlayer exchange calculation. The $n=6$ correction with nonzero expectation values are
\begin{align}
C^{\perp}_{ 6}
=&6!\int_\tau d(6\dots1)\Big\{
\langle\mathcal{H}_t(\tau_6)\mathcal{H}_{\perp}(\tau_5)\mathcal{H}^\dagger_t(\tau_4)\mathcal{H}_t(\tau_3)\mathcal{H}_{\perp}(\tau_2)]\mathcal{H}^{\dagger}_{t}(\tau_1)|\tilde{\Psi}\rangle_c& \leftarrow C^{\perp}_{6,SE1}
\notag\\
&+\langle\mathcal{H}_t(\tau_6)[\mathcal{H}_{\perp}(\tau_5)\mathcal{H}_t(\tau_4)+\mathcal{H}_t(\tau_5)\mathcal{H}_{\perp}(\tau_4)][\mathcal{H}_{\perp}(\tau_3)\mathcal{H}_t^\dagger(\tau_2)+\mathcal{H}_t^\dagger(\tau_3)\mathcal{H}_{\perp}(\tau_2)]\mathcal{H}^{\dagger}_{t}(\tau_1)|\tilde{\Psi}\rangle_c \Big\},&\leftarrow C^{\perp}_{6,SE2}
\end{align}
where the subscript $c$ means that we keep only processes which have a \emph{connected} and \emph{close} hopping path on the lattice.
In the above, the first line and the second line correspond to the SE1 and SE2 contributions. Integrating out the imaginary time and keeping only the linear term in $\beta$ (low-temperature limit), we obtain
\begin{align}
C^{\perp}_{6,SE1}
=
&6!\int_\tau d(6\dots1)
\chi^{\ell_6}_{\dot{\mathbf{r}}_6\sigma_6}\bar{\chi}^{\ell_5}_{\dot{\mathbf{r}}_5\sigma_6}\bar{\chi}^{\ell_1}_{\dot{\mathbf{r}}_1\sigma_1}\chi^{\ell_2}_{\dot{\mathbf{r}}_2\sigma_1}t_{ 6\dot{6}}(\tau_6)T^{\ell_6\ell_5}_{ 6 5}(\tau_5)\bar{t}_{ 5 \dot{ 5}}(\tau_4)
t_{\dot{1} 1} (\tau_1)T^{\ell_1\ell_2}_{ 1 2}(\tau_2) \bar{t}_{ 2 \dot{ 2}}(\tau_3)\langle \tilde{\Psi}|\mathsf{d}^{\ell_6}_{\dot{6}}\mathsf{d}^{\ell_5\dagger}_{ \dot{5}}\mathsf{d}^{\ell_2}_{ \dot{2}} \mathsf{d}^{\ell_1\dagger}_{\dot{1}}|\tilde{\Psi}\rangle\notag\\
=
&
6!\beta\frac{1}{2}(1+\mathbf{s}^{\ell_1}_{\dot{\mathbf{r}}_1}\mathbf{s}^{\ell_2}_{\dot{\mathbf{r}}_2})\frac{P_{\dot{\alpha}_2}t_{ 4\dot{1}}T^{\ell_2\ell_1}_{ 34 } t_{ \dot{ 2}3 }
t_{\dot{1} 1}T^{\ell_1\ell_2}_{ 1 2} t_{ 2 \dot{ 2}}}{\mathcal{E}_{\dot{1}1}\mathcal{E}_{\dot{1}2}\mathcal{E}_{\dot{1}3}\mathcal{E}_{\dot{1}4}(\omega_{\dot{1}}-\bar{\omega}_{\dot{2}})}.
\end{align}
Turning to the SE2 process, we have
\begin{align}
C^{\perp}_{6,SE2}
=
&6!\int_\tau d(6\dots1)
X^{\sigma_6 \sigma_5}_{\ell_6\dot{\mathbf{r}}_6,\ell_5\dot{\mathbf{r}}_5,\eta'}\bar{X}^{\sigma_1 \sigma_2}_{\ell_1\dot{\mathbf{r}}_1,\ell_2\dot{\mathbf{r}}_2,\eta}\Big\{
\Xi_{-1}(\tau_1|_{\dot{1}1},\tau_2|_{\dot{2}2})T^{\ell_1\ell_2}_{ 13}(\tau_3)
+
\eta
t_{\dot{1} 1}(\tau_1)T^{\ell_1\ell_2}_{1 2}(\tau_2) t_{ \dot{2} 3}(\tau_3)\Big\}\mathrm{e}^{\tau_3\Theta^{\eta}_{23}-\tau_4\Theta^{\eta'}_{54}}\notag\\
&\Big\{
\Xi_{-1}(\tau_6|_{6\dot{6}},\tau_5|_{5\dot{5}})T^{\ell_6\ell_5}_{ 64}(\tau_4)
+
\eta'
t_{ 6\dot{6}}(\tau_6)T^{\ell_6\ell_5}_{65 }(\tau_5) t_{ 4 \dot{5}}(\tau_4)\Big\}
\langle\tilde{\Psi}|\mathsf{d}^{\ell_6}_{\dot{ 6}} \mathsf{d}^{\ell_5}_{\dot{5}}p^{\ell_5,\dagger}_{ 5 \sigma_5}p^{\ell_5,\dagger}_{ 4 \sigma_6}
p^{\ell_2}_{ 3 \sigma_1}p^{\ell_2}_{ 2 \sigma_2} \mathsf{d}^{\ell_2\dagger}_{ \dot{2}}\mathsf{d}^{\ell_1\dagger}_{ \dot{1}}|\tilde{\Psi}\rangle\notag\\
=
&
6!\beta\eta\frac{1}{2}(1+\mathbf{s}^{\ell_1}_{\dot{\mathbf{r}}_1}\mathbf{s}^{\ell_2}_{\dot{\mathbf{r}}_2})\frac{1}{2}\frac{t_{ 2\dot{2}} t_{ 6\dot{1}} T^{\ell_1\ell_2}_{ 63}((\mathcal{E}_{\dot{1}3})^2+(\mathcal{E}_{\dot{2}2})^2)}{\mathcal{E}_{\dot{1}1}\mathcal{E}_{\dot{1}6}(\mathcal{E}_{\dot{1}3}+\mathcal{E}_{\dot{2}2}+\Theta^\eta_{23})(\mathcal{E}_{\dot{1}3}\mathcal{E}_{\dot{2}2})^2}
\Big[t_{\dot{2} 2}t_{\dot{1} 1} T^{\ell_1\ell_2}_{ 13}
-\eta t_{\dot{2} 3}t_{\dot{1} 1} T^{\ell_1\ell_2}_{ 12}\Big].
\end{align}
Therefore, combining the SE1 and SE2 combination, we have
\begin{equation}
C^\perp_6=C^{\perp}_{6,SE1}+C^{\perp}_{6,SE2}.
\end{equation}
The spin-dependent terms in $C^\perp_6$ gives the $\mathcal{J}^{\perp}$ in the main text.

\subsection{Interlayer DM interaction}

The nonzero terms in the $n=7$ correction are
\begin{align}
C^{\bm{\mathcal{D}}^{\perp}}_7=&-7!\int_\tau d(7\dots1)\Big\{2\mathrm{Re}\Big[\langle\tilde{\Psi}|\mathcal{H}_t(\tau_7)\mathcal{H}_{\perp}(\tau_6)\mathcal{H}^\dagger_t(\tau_5)\mathcal{H}_t(\tau_4)[\mathcal{H}_{\perp}(\tau_3)\mathcal{V}_{\lambda}(\tau_2)+\mathcal{V}_{\lambda}(\tau_3)\mathcal{H}_{\perp}(\tau_2)]\mathcal{H}^{\dagger}_{t}(\tau_1)|\tilde{\Psi}\rangle_c\Big]&\leftarrow C^{\bm{\mathcal{D}}^{\perp}}_{7,SE1}\;\,
\notag\\
&+\langle \tilde{\Psi}|\mathcal{H}_t(\tau_7)[ \mathcal{H}_{\perp}(\tau_6)\mathcal{H}_t(\tau_5)+\mathcal{H}_t(\tau_6) \mathcal{H}_{\perp}(\tau_5)]\mathcal{V}_{\lambda}(\tau_4)[ \mathcal{H}_{\perp}(\tau_3)\mathcal{H}^\dagger_t(\tau_2)+\mathcal{H}^\dagger_t(\tau_3) \mathcal{H}_{\perp}(\tau_2)]\mathcal{H}^\dagger_t(\tau_1)|\tilde{\Psi}\rangle_c&\leftarrow C^{\bm{\mathcal{D}}^{\perp}}_{7,SE2a}\notag\\
&+2\mathrm{Re}\Big[\langle \tilde{\Psi}|\mathcal{H}_t(\tau_7)[ \mathcal{H}_{\perp}(\tau_6)\mathcal{H}_t(\tau_5)+\mathcal{H}_t(\tau_6) \mathcal{H}_{\perp}(\tau_5)] [\mathcal{H}_{\perp}(\tau_4)\mathcal{V}_{\lambda}(\tau_3)\mathcal{H}_t^\dagger(\tau_2)+ \mathcal{H}_t^\dagger(\tau_4) (\mathcal{H}_{\perp}(\tau_3)\mathcal{V}_{\lambda}(\tau_2)&\leftarrow C^{\bm{\mathcal{D}}^{\perp}}_{7,SE2b}\notag\\
&\quad\quad\quad+\mathcal{V}_{\lambda}(\tau_2)\mathcal{H}_{\perp}(\tau_2))]\mathcal{H}_t^\dagger(\tau_1)\Big]|\tilde{\Psi}\rangle_c\Big]\Big\}.
\end{align}
Similar to the interlayer Heisenberg exchange, we calculate the contribution from SE1 (first line) and SE2 (last three lines) separately. For SE1, the calculation is straightforward which is
\begin{align}
C^{\bm{D}^{\perp}}_{7,SE1}=&-7!\int_{\tau}d(7\dots1)t^{m_6}_{ 6\dot{6}} (\tau_7)T^{\ell_5\ell_6}_{ 56}2\mathrm{Re}\Big[(\tau_6) \bar{t}_{ \dot{ 5}5 }(\tau_5)t_{\dot{1}1}(\tau_1) \Big[T^{\ell_1\ell_2}_{12}(\tau_2)\Lambda_{23}^{j}(\tau_3)+\Lambda_{12}^{j}(\tau_2)T^{\ell_{1}\ell_2}_{23}(\tau_3)\Big]t_{3\dot{2}}(\tau_4)\Big] \notag\\
&\bar{\chi}^{\ell_6}_{\dot{\mathbf{r}}_6\sigma_6}\chi^{\ell_5}_{\dot{\mathbf{r}}_5\sigma_6}\bar{\slashed{\chi}}^{\ell_1 j}_{ \dot{\mathbf{r}}_1\sigma_3}\chi^{\ell_2}_{\mathbf{r}_2\sigma_3}\langle\tilde{\Psi}|\mathsf{d}^{\ell_6}_{\dot{6}}\mathsf{d}^{\ell_5}_{ \dot{5}}
\mathsf{d}^{\ell_2}_{\dot{2}+} \mathsf{d}^{\ell_1\dagger}_{\dot{1} }|\tilde{\Psi}\rangle\notag\\
=&-7!\beta\mathbf{s}^{\ell_1}_{\dot{\mathbf{r}}_1}\times\mathbf{s}^{\ell_2}_{\dot{\mathbf{r}}_2}\frac{
t_{\dot{1} 6} T^{\ell_2\ell_1}_{ 5 6} t_{ 5 \dot{ 2}}
t_{\dot{1}1} \Big(T^{\ell_1\ell_2}_{12}i\Lambda_{23}^{j}+i\Lambda_{12}^{j}T^{\ell_{1}\ell_2}_{23}\Big)t_{3\dot{2}}}{(\omega_{\dot{1}}-\bar{\omega}_{\dot{2}})\mathcal{E}_{\dot{1}1}\mathcal{E}_{\dot{1}2}\mathcal{E}_{\dot{1}3}\mathcal{E}_{\dot{1}5}\mathcal{E}_{\dot{1}6}}.
\end{align}
The calculation for the SE2 contribution is tedious. We break the calculation into two parts. The first part is
\begin{align}
C^{\bm{D}^{\perp}}_{7,SE2a}&=7!\int_\tau d(7\dots1)
\Big\{
\Xi_{-1}(\tau_7|_{7\dot{7}},\tau_6|_{6\dot{6}})
T^{\ell_7\ell_6}_{75}(\tau_5)+\xi
t_{7\dot{7} }^{ m_7}(\tau_7)T^{\ell_7\ell_6}_{ 76}(\tau_6) t^{m_6}_{ 5\dot{6} }(\tau_5) \Big\}\mathrm{e}^{\tau_5\Theta^{\xi}_{65}}\mathrm{e}^{(\tau_3-\tau_4)\Theta^{\eta }_{23}}\Lambda_{34}^{j}(\tau_4)\mathrm{e}^{\tau_4\Theta^{\eta'}_{24}}\notag\\
&
\Big\{[\Xi_{-1}(\tau_1|_{\dot{1}1},\tau_2|_{\dot{2}2})
T^{\ell_1\ell_2}_{ 13}(\tau_3)- t_{\dot{1} 1}^{ m_1}(\tau_1)T^{\ell_1\ell_2}_{1 3}(\tau_2)t_{ \dot{2} 2}(\tau_3) ]-\eta[
\Xi_{-1}(\tau_1|_{\dot{1}1},\tau_2|_{\dot{2}3})
T^{\ell_1\ell_2}_{ 12}(\tau_3)-t_{\dot{1} 1}^{ m_1}(\tau_1)T^{\ell_1\ell_2}_{1 2}(\tau_2) t_{ \dot{2} 3}(\tau_3)]\Big\}\notag\\
&
(\bar{\slashed{X}}^{j,\sigma_1 \sigma_2}_{\ell_1\dot{\mathbf{r}}_1,\ell_2\dot{\mathbf{r}}_2,\eta}+\eta'\bar{\slashed{X}}^{j,\sigma_2 \sigma_1}_{\ell_1\dot{\mathbf{r}}_1,\ell_2\dot{\mathbf{r}}_2,\eta})X^{\sigma_7\sigma_6}_{\ell_7\dot{7}\ell_6\dot{6},\xi}\langle\tilde{\Psi}|\mathsf{d}^{\ell_7}_{ \dot{7}}\mathsf{d}^{\ell_6}_{\dot{ 6}}p^{\ell_6\dagger}_{ 6 \sigma_6}p^{\ell_6\dagger}_{ 5 \sigma_7}p^{\ell_2}_{4 \sigma_1}p^{\ell_2}_{ 2\sigma_2}
\mathsf{d}^{\ell_2\dagger}_{ \dot{2}}\mathsf{d}^{\ell_1\dagger}_{\dot{1}}|\tilde{\Psi}\rangle.
\end{align}
Using the following spin-sum result,
\begin{align}
2\bar{\slashed{X}}^{j,\sigma_1 \sigma_2}_{\ell_1\dot{\mathbf{r}}_1,\ell_2\dot{\mathbf{r}}_2,\eta}X^{\sigma_1\sigma_2}_{\ell_1\dot{\mathbf{r}}_1,\ell_2\dot{\mathbf{r}}_2,\xi}=&\frac{1}{2}\Big[ \mathbf{s}_{\dot{\mathbf{r}}_1}+\eta\xi \mathbf{s}_{\dot{\mathbf{r}}_2}+\frac{\xi+\eta}{2}(-\mathbf{s}_{\dot{\mathbf{r}}_1}- \mathbf{s}_{\dot{\mathbf{r}}_2})+\frac{\eta-\xi}{2}i\mathbf{s}_{\dot{\mathbf{r}}_1}\times\mathbf{s}_{\dot{\mathbf{r}}_2}\Big],
\end{align}
and integrating out $\tau$, this gives the result for the first part as
\begin{align}
C^{\bm{D}^{\perp}}_{7,SE2a}=&7!\beta\frac{i\mathbf{s}^{\ell_1}_{\dot{\mathbf{r}}_1}\times\mathbf{s}^{\ell_2}_{\dot{\mathbf{r}}_2}t_{7\dot{1}}t_{\dot{1}1}\Lambda^j_{34}}{\mathcal{E}_{\dot{1}1}\mathcal{E}_{\dot{1}7}(\mathcal{E}_{\dot{1}2}+\mathcal{E}_{\dot{2}3}+\Theta_{23}^{\eta})(\mathcal{E}_{\dot{1}2}+\mathcal{E}_{\dot{2}4}+\Theta_{24}^{\xi})}\frac{1}{2}\Big[T^{\ell_1\ell_2}_{13}t_{\dot{2}2}\Big(\frac{1}{\mathcal{E}_{\dot{1}3}}-\frac{1}{\mathcal{E}_{\dot{2}2}}\Big)-\eta T^{\ell_1\ell_2}_{12}t_{\dot{2}3}\Big(\frac{1}{\mathcal{E}_{\dot{2}3}}-\frac{1}{\mathcal{E}_{\dot{1}2}}\Big)\Big]\notag\\
&
\Big[T^{\ell_2\ell_1}_{47}t_{2\dot{2}}\Big(\frac{1}{\mathcal{E}_{\dot{2}2}}-\frac{1}{\mathcal{E}_{\dot{1}4}}\Big)+\xi T^{\ell_2\ell_1}_{27}t_{4\dot{2}}\Big(\frac{1}{\mathcal{E}_{\dot{2}4}}-\frac{1}{\mathcal{E}_{\dot{1}2}}\Big)\Big].
\end{align}
Turning to the second part, we have
\begin{align}
C^{\bm{D}^{\perp}}_{7,SE2b}
=&-7!\int_\tau d(7\dots1)
\Big\{
\Xi_{-1}(\tau_7|_{7\dot{7}},\tau_6|_{6\dot{6}})
T^{\ell_6\ell_7}_{57}(\tau_5)+\xi
t_{7\dot{7} }^{ m_7}(\tau_7)T^{\ell_6\ell_7}_{ 67}(\tau_6) t^{m_6}_{ 5\dot{6} }(\tau_5) \Big\}
\notag\\
&
\mathrm{e}^{\tau_5\Theta^{\xi }_{65}}\Big\{\Big[
(\bar{\chi}^{\sigma_1}_{\ell_1\dot{\mathbf{r}}_1\dot{\sigma}_1}\bar{\slashed{\chi}}^{j,\sigma_2}_{\ell_2\dot{\mathbf{r}}_2\dot{\sigma}_2}+\eta\bar{\chi}^{\sigma_2}_{\ell_1\dot{\mathbf{r}}_1\dot{\sigma}_1}\bar{\slashed{\chi}}^{j,\sigma_1}_{\ell_2\dot{\mathbf{r}}_2\dot{\sigma}_2})\Xi_{-1}(\tau_1|_{\dot{1}1},\tau_2|_{\dot{2}3})
\Lambda^{j}_{32}(\tau_3)\notag\\
&
+\eta(\bar{\chi}^{\sigma_1}_{\ell_2\dot{\mathbf{r}}_2\dot{\sigma}_2}\bar{\slashed{\chi}}^{j,\sigma_2}_{\ell_1\dot{\mathbf{r}}_1\dot{\sigma}_1}+\eta\bar{\chi}^{\sigma_2}_{\ell_2\dot{\mathbf{r}}_2\dot{\sigma}_2}\bar{\slashed{\chi}}^{j,\sigma_1}_{\ell_1\dot{\mathbf{r}}_1\dot{\sigma}_1})\Xi_{-1}(\tau_1|_{\dot{2}2},\tau_2|_{\dot{1}3})
\Lambda^{j}_{31}(\tau_3)
\Big]
T^{\ell_1\ell_2}_{14}(\tau_4)
\notag\\
&
-
(\bar{\slashed{\chi}}^{\sigma_2}_{\dot{\mathbf{r}}_1\dot{\sigma}_1}\bar{\chi}^{\sigma_1}_{\dot{\mathbf{r}}_2\dot{\sigma}_2}+\eta \bar{\slashed{\chi}}^{\sigma_1}_{\dot{\mathbf{r}}_1\dot{\sigma}_1}\bar{\chi}^{\sigma_2}_{\dot{\mathbf{r}}_2\dot{\sigma}_2})
t_{\dot{1}1}(\tau_1) \Big[T^{\ell_1\ell_2}_{13}(\tau_2)\Lambda_{32}^{j}(\tau_3)+\Lambda_{13}^{j}(\tau_2)T^{\ell_{1}\ell_2}_{32}(\tau_3)\Big]t_{\dot{2}4}(\tau_4)
\Big\}\mathrm{e}^{\tau_4\Theta^\eta_{24}}\notag\\
&X^{\sigma_7\sigma_6}_{\ell_7\dot{7}\ell_6\dot{6},\xi}\langle\tilde{\Psi}|\mathsf{d}^{\ell_7}_{ \dot{7}}\mathsf{d}^{\ell_6}_{\dot{ 6}}p^{\ell_6\dagger}_{ 6 \sigma_6}p^{\ell_6\dagger}_{ 5 \sigma_7}
p^{\ell_2}_{ 4\sigma_1}p^{\ell_2}_{2\sigma_2}
\mathsf{d}^{\ell_2\dagger}_{ \dot{2}}\mathsf{d}^{\ell_1\dagger}_{\dot{1}}|\tilde{\Psi}\rangle.
\end{align}
Keeping only the real terms, we obtain
\begin{align}
C^{\bm{D}^{\perp}}_{7,SE2b}=&-7!\beta
\frac{i\mathbf{s}^{\ell_1}_{\dot{\mathbf{r}}_1}\times\mathbf{s}^{\ell_2}_{\dot{\mathbf{r}}_2}}{(\mathcal{E}_{\dot{1}1}+\mathcal{E}_{\dot{2}2})(\mathcal{E}_{\dot{1}2}+\mathcal{E}_{\dot{2}4}+\Theta_{24}^{\xi})}\frac{t_{7\dot{1}}}{\mathcal{E}_{\dot{1}7}}\Big[t_{4\dot{2}}T_{27}^{\ell_2\ell_1}\Big(\frac{1}{\mathcal{E}_{\dot{2}4}}-\frac{1}{\mathcal{E}_{\dot{1}2}}\Big)+\xi t_{2\dot{2}}T_{47}^{\ell_2\ell_1}\Big(\frac{1}{\mathcal{E}_{\dot{2}2}}-\frac{1}{\mathcal{E}_{\dot{1}4}}\Big)\Big]\notag\\
&
\Big[\frac{T^{\ell_1\ell_2}_{14}t_{\dot{1}1}t_{\dot{2}3}\Lambda^{j}_{32}}{\mathcal{E}_{\dot{1}1}\mathcal{E}_{\dot{2}3}}-\frac{T^{\ell_1\ell_2}_{14}t_{\dot{1}3}t_{\dot{2}2}\Lambda^{j}_{31}}{\mathcal{E}_{\dot{2}2}\mathcal{E}_{\dot{1}3}}
-\xi \frac{(T^{\ell_1\ell_2}_{13}\Lambda^j_{32}+\Lambda^j_{13}T^{\ell_1\ell_2}_{32})t_{\dot{1}1}t_{\dot{2}4}(\mathcal{E}_{\dot{1}1}+\mathcal{E}_{\dot{2}2})}{\mathcal{E}_{\dot{1}1}\mathcal{E}_{\dot{1}2}\mathcal{E}_{\dot{1}3}}\Big].
\end{align}
Combining all the contributions, 
\begin{equation}
C^{\bm{\mathcal{D}}^{\perp}}_7=C^{\bm{\mathcal{D}}^{\perp}}_{7,SE1}+C^{\bm{\mathcal{D}}^{\perp}}_{7,SE2a}+C^{\bm{\mathcal{D}}^{\perp}}_{7,SE2b}.
\end{equation}
The above leads to the interlayer DM interaction $\bm{\mathcal{D}}^{\perp}$ in the main text. The calculation of the next-nearest DM interaction $\bm{\mathcal{D}}'$ is the same as $\bm{\mathcal{D}}^{\perp}$ which can be obtained by replacing the interlayer hopping $T^{\ell_1\ell_"}_{12}\to v_{12}$.

\subsection{Intralayer symmetric exchange tensor}

The procedure for evaluating intralayer symmetric exchange tensor is also the same as the previous calculations. In the derivation, we omit the layer index for simplicity since the calculation only involves one layer. This exchange coupling also arises from the $n=6$ correction [see Eq. \eqref{eqn:C6}]
\begin{align}
C^{\mathcal{K}}_6=&6!\int d(654321)\Big\{\langle\tilde{\Psi}|\mathcal{H}_{t}(\tau_4)\mathcal{V}_\lambda(\tau_3)\mathcal{H}^\dagger_{t}(\tau_4)\mathcal{H}_{t}(\tau_3)\mathcal{V}_\lambda(\tau_2)\mathcal{H}^\dagger_t(\tau_1)|\tilde{\Psi}\rangle_c
&\leftarrow C^{\mathcal{K}}_{6,SE1}
\notag\\
&+\langle\tilde{\Psi}|\mathcal{H}_t(\tau_6)[\mathcal{V}_\lambda(\tau_5)\mathcal{H}_t(\tau_4)+
\mathcal{H}_t(\tau_5)\mathcal{V}_\lambda(\tau_4)][\mathcal{V}_\lambda(\tau_3)\mathcal{H}^\dagger_t(\tau_2)+
\mathcal{H}^\dagger_t(\tau_3)\mathcal{V}_\lambda(\tau_2)]\mathcal{H}^\dagger_t(\tau_1)|\tilde{\Psi}\rangle_c\Big\}.
&\leftarrow C^{\mathcal{K}}_{6,SE1}
\end{align}
Similarly, we first calculate the contribution from SE1. The nonzero 6th order process that contribute to $\mathcal{K}^{jj'}$ is
\begin{align}
C_{\mathcal{K},6}^{\text{SE1}}
=&
6!\int_\tau d(6\dots1)P_{ \dot{\alpha}_3} t_{\dot{1} 1}(\tau_1)\Lambda_{\mathbf{r}_1,j}^{\alpha_1 \alpha_2}(\tau_2)\delta_{\mathbf{r}_1 \mathbf{r}_2}\bar{t}_{2\dot{3}}(\tau_3)
 \bar{t}_{ \dot{3}6}(\tau_4)\Lambda_{\mathbf{r}_6j'}^{\alpha_5 \alpha_6}(\tau_5)\delta_{\mathbf{r}_5\mathbf{r}_6} t_{6 \dot{1}}(\tau_6) \bar{\chi}_{ \dot{\mathbf{r}}_3\sigma_5}\tau^{j'}_{\sigma_5 \sigma_6}\chi_{ \dot{\mathbf{r}}_1\sigma_6}\chi_{ \dot{\mathbf{r}}_3\sigma_2}\tau^{j}_{\sigma_1 \sigma_2}\bar{\chi}_{ \dot{\mathbf{r}}_1\sigma_1}\notag\\
=&6!\beta\frac{(s_{\dot{\mathbf{r}}_1j}s_{\dot{\mathbf{r}}_2j'}+s_{\dot{\mathbf{r}}_1j'}s_{\dot{\mathbf{r}}_2j})}{2}
\frac{P_{ \dot{\alpha}_2}\Lambda_{\mathbf{r}_1j}^{\alpha_1 \alpha_2}\delta_{\mathbf{r}_1 \mathbf{r}_2}\Lambda_{\mathbf{r}_5 j'}^{\alpha_5 \alpha_6}\delta_{\mathbf{r}_5 \mathbf{r}_6} t_{\dot{2}  5} t_{6 \dot{1} } t_{ 2\dot{2}}   t_{ \dot{1} 1}}{\mathcal{E}_{ \dot{1} 1}\mathcal{E}_{ \dot{1} 2}\mathcal{E}_{ \dot{1} 5}\mathcal{E}_{ \dot{1} 6}(\omega_{\dot{1}}-\bar{\omega}_{ \dot{2}})}
\end{align}
Then, the SE2 process is
\begin{align*}
C^{SE2}_{\mathcal{K},6}&
=6!\int_\tau d(6\dots1)\langle\tilde{\Psi}|\mathsf{d}_{\dot{6}}\mathsf{d}_{\dot{5}}p^\dagger_{ 6\sigma_6}p^\dagger_{ 4\sigma_4}
p_{ 3\sigma_3}p_{ 1\sigma_1}
\mathsf{d}^{ \dagger}_{\dot{2}}\mathsf{d}^{\dagger}_{\dot{1} \dot{\sigma}_1}|\tilde{\Psi}\rangle\frac{1}{4}\mathrm{e}^{-\tau_4\Theta^{\xi'}_{64}}\mathrm{e}^{\tau_3\Theta^{\eta'}_{13}}
\notag\\
&\Big\{\Xi_{\xi}(\tau_6|_{6\dot{6}},\tau_5|_{5\dot{5}})\mathrm{e}^{(\tau_5-\tau_4)\Theta^{\xi }_{65}}\Lambda_{54}^{j}(\tau_4)
(\bar{\slashed{X}}^{j',\sigma_4\sigma_6}_{\ell_6\dot{\mathbf{r}}_6,\ell_5\dot{\mathbf{r}}_5,\xi}\!+\!\xi'\bar{\slashed{X}}^{j',\sigma_6\sigma_4}_{\ell_6\dot{\mathbf{r}}_6,\ell_5\dot{\mathbf{r}}_5,\xi})\!
+\!
(\bar{\slashed{\chi}}^{j',\sigma_6}_{ \dot{\mathbf{r}}_6\dot{\sigma}_6}\bar{\chi}^{\sigma_4}_{\dot{\mathbf{r}}_5\dot{\sigma}_5}\!+\!\xi'\bar{\slashed{\chi}}^{j',\sigma_4}_{ \dot{\mathbf{r}}_6\dot{\sigma}_6}\bar{\chi}^{\sigma_6}_{\dot{\mathbf{r}}_5\dot{\sigma}_5})t_{\dot{6}  5} (\tau_6) \Lambda_{56}^{j'}(\tau_5)t_{ \dot{5}4}(\tau_4)
\Big\}\notag\\
&\Big\{\Xi_{\eta}(\tau_1|_{\dot{1}1},\tau_2|_{\dot{2}2})\mathrm{e}^{(\tau_2-\tau_3)\Theta^{\eta }_{12}}\Lambda_{23}^{j}(\tau_3)
(\bar{\slashed{X}}^{j,\sigma_3\sigma_1}_{\ell_1\dot{\mathbf{r}}_1,\ell_2\dot{\mathbf{r}}_2,\eta}\!+\!\eta'\bar{\slashed{X}}^{j,\sigma_1\sigma_3}_{\ell_1\dot{\mathbf{r}}_1,\ell_2\dot{\mathbf{r}}_2,\eta})
\!+\!
(\bar{\slashed{\chi}}^{j,\sigma_1}_{ \dot{\mathbf{r}}_1\dot{\sigma}_1}\bar{\chi}^{\sigma_3}_{\dot{\mathbf{r}}_2\dot{\sigma}_2}\!+\!\eta'\bar{\slashed{\chi}}^{j,\sigma_3}_{ \dot{\mathbf{r}}_1\dot{\sigma}_1}\bar{\chi}^{\sigma_1}_{\dot{\mathbf{r}}_2\dot{\sigma}_2})t_{\dot{1}  2} (\tau_1) \Lambda_{21}^{j}(\tau_2)t_{ \dot{2}3}(\tau_3)
\Big\}.
\end{align*}
Integrating out $\tau$ and keeping only the $\mathbf{s}_{\dot{\mathbf{r}}}$-dependent terms, we obtain
\begin{align}
C^{SE2}_{\mathcal{K},6}=&6!\beta\frac{\frac{1}{4}(s_{\dot{\mathbf{r}_1}j'}s_{\dot{\mathbf{r}_2}j}+s_{\dot{\mathbf{r}_1}j}s_{\dot{\mathbf{r}_2}j'})}{(\mathcal{E}_{\dot{1}1}+\mathcal{E}_{\dot{2}3}+\Theta_{13}^{\xi'})}\Big\{
\Big[-\xi' \frac{\Lambda_{\mathbf{r}_3j}^{\alpha_2\alpha_3}\delta_{\mathbf{r}_2\mathbf{r}_3} t_{\dot{1}1}t_{\dot{2}2}}{(\mathcal{E}_{\dot{1}1}+\mathcal{E}_{\dot{2}2}+\Theta_{12}^{-1})}\Big(\frac{1}{\mathcal{E}_{\dot{1}1}}-\frac{\xi}{\mathcal{E}_{\dot{2}2}}\Big)+
\frac{\Lambda_{\mathbf{r}_1j}^{\alpha_2 \alpha_1}\delta_{\mathbf{r}_1\mathbf{r}_2} t_{\dot{1}2}t_{\dot{2}3}}{\mathcal{E}_{\dot{1}1}\mathcal{E}_{\dot{1}2}}\Big]\notag\\
&\Big[\frac{\tfrac{1+\xi}{2}\Lambda_{\mathbf{r}_1j'}^{\alpha_5\alpha_1}\delta_{\mathbf{r}_5\mathbf{r}_1} }{(\mathcal{E}_{\dot{1}3}+\mathcal{E}_{\dot{2}5}+\Theta_{35}^{\xi})}\Big(\frac{t_{3\dot{2}}t_{5\dot{1}}}{\mathcal{E}_{\dot{1}5}}-\frac{\xi t_{3\dot{2}}t_{5\dot{1}}}{\mathcal{E}_{\dot{2}3}}
+
\frac{t_{3\dot{1}}t_{5\dot{2}}}{\mathcal{E}_{\dot{1}3}}-\frac{\xi t_{3\dot{1}}t_{5\dot{2}}}{\mathcal{E}_{\dot{2}5}}\Big)
+
\xi'\Lambda_{\mathbf{r}_1j'}^{\alpha_5\alpha_1}\delta_{\mathbf{r}_5\mathbf{r}_1} \Big(\frac{t_{3\dot{2}}t_{5\dot{1}}}{\mathcal{E}_{\dot{1}1}\mathcal{E}_{\dot{1}5}}
-
\frac{t_{3\dot{1}}t_{5\dot{2}}}{\mathcal{E}_{\dot{2}1}\mathcal{E}_{\dot{2}5}}
\Big)
\notag\\
&
-
\frac{\tfrac{1+\xi}{2}\xi'\Lambda_{\mathbf{r}_1j'}^{\alpha_5\alpha_3}\delta_{\mathbf{r}_5\mathbf{r}_3} }{(\mathcal{E}_{\dot{1}1}+\mathcal{E}_{\dot{2}5}+\Theta_{15}^{\xi})}\Big(\frac{t_{1\dot{2}}t_{5\dot{1}}}{\mathcal{E}_{\dot{1}5}}-\frac{\xi t_{1\dot{2}}t_{5\dot{1}}}{\mathcal{E}_{\dot{2}1}}
+
\frac{t_{1\dot{1}}t_{5\dot{2}}}{\mathcal{E}_{\dot{1}1}}-\frac{\xi t_{1\dot{1}}t_{5\dot{2}}}{\mathcal{E}_{\dot{2}5}}\Big)
-\Lambda_{\mathbf{r}_1j'}^{\alpha_5\alpha_3}\delta_{\mathbf{r}_5\mathbf{r}_3} \Big(
\frac{t_{1\dot{2}}t_{5\dot{1}}}{\mathcal{E}_{\dot{1}3}\mathcal{E}_{\dot{1}5}}
-
\frac{t_{1\dot{1}}t_{5\dot{2}}}{\mathcal{E}_{\dot{2}3}\mathcal{E}_{\dot{2}5}}
\Big)
\Big]
\Big\}.
\end{align}
We note that this 6th-order correction also includes the isotropic Heisenberg exchange [see the spin-sum in Eq. \eqref{eqn:tensor-spin}]. It will contribute to $\mathcal{J}^{\perp}$. However, this correction is small as compared to the 4th-order correction. Therefore, we discard them. Combining with the result in SE1, we then arrive at the intralayer symmetric exchange
\begin{align}
\mathcal{K}^{jj'}_{\dot{\mathbf{r}}_1\dot{\mathbf{r}}_2}=&\frac{2^2}{3^2}\frac{1}{2}\Big\{\frac{P_{ \dot{\alpha}_2}\Lambda_{\mathbf{r}_1j}^{\alpha_1 \alpha_2}\delta_{\mathbf{r}_1 \mathbf{r}_2}\Lambda_{\mathbf{r}_5 j'}^{\alpha_5 \alpha_6}\delta_{\mathbf{r}_5 \mathbf{r}_6} t_{\dot{2}  5} t_{6 \dot{1} } t_{ 2\dot{2}}   t_{ \dot{1} 1}}{\mathcal{E}_{ \dot{1} 1}\mathcal{E}_{ \dot{1} 2}\mathcal{E}_{ \dot{1} 5}\mathcal{E}_{ \dot{1} 6}(\omega_{\dot{1}}-\bar{\omega}_{ \dot{2}})}
-
\frac{1/2}{(\mathcal{E}_{\dot{1}1}+\mathcal{E}_{\dot{2}3}+\Theta_{13}^{\xi'})}
\Big[\xi' \frac{\Lambda_{\mathbf{r}_3j}^{\alpha_2\alpha_3}\delta_{\mathbf{r}_2\mathbf{r}_3} t_{\dot{1}1}t_{\dot{2}2}}{(\mathcal{E}_{\dot{1}1}+\mathcal{E}_{\dot{2}2}+\Theta_{12}^{-1})}\Big(\frac{1}{\mathcal{E}_{\dot{1}1}}-\frac{\xi}{\mathcal{E}_{\dot{2}2}}\Big)
\notag\\
&
-
\frac{\Lambda_{\mathbf{r}_1j}^{\alpha_2 \alpha_1}\delta_{\mathbf{r}_1\mathbf{r}_2} t_{\dot{1}2}t_{\dot{2}3}}{\mathcal{E}_{\dot{1}1}\mathcal{E}_{\dot{1}2}}\Big]
\Big[\frac{\tfrac{1+\xi}{2}\Lambda_{\mathbf{r}_1j'}^{\alpha_5\alpha_1}\delta_{\mathbf{r}_5\mathbf{r}_1} }{(\mathcal{E}_{\dot{1}3}+\mathcal{E}_{\dot{2}5}+\Theta_{35}^{\xi})}\Big(\frac{t_{3\dot{2}}t_{5\dot{1}}}{\mathcal{E}_{\dot{1}5}}-\frac{\xi t_{3\dot{2}}t_{5\dot{1}}}{\mathcal{E}_{\dot{2}3}}
+
\frac{t_{3\dot{1}}t_{5\dot{2}}}{\mathcal{E}_{\dot{1}3}}-\frac{\xi t_{3\dot{1}}t_{5\dot{2}}}{\mathcal{E}_{\dot{2}5}}\Big)
+
\xi'\Lambda_{\mathbf{r}_1j'}^{\alpha_5\alpha_1}\delta_{\mathbf{r}_5\mathbf{r}_1} \Big(\frac{t_{3\dot{2}}t_{5\dot{1}}}{\mathcal{E}_{\dot{1}1}\mathcal{E}_{\dot{1}5}}
-
\frac{t_{3\dot{1}}t_{5\dot{2}}}{\mathcal{E}_{\dot{2}1}\mathcal{E}_{\dot{2}5}}
\Big)
\notag\\
&
-
\frac{\tfrac{1+\xi}{2}\xi'\Lambda_{\mathbf{r}_3j'}^{\alpha_5\alpha_3}\delta_{\mathbf{r}_5\mathbf{r}_3} }{(\mathcal{E}_{\dot{1}1}+\mathcal{E}_{\dot{2}5}+\Theta_{15}^{\xi})}\Big(\frac{t_{1\dot{2}}t_{5\dot{1}}}{\mathcal{E}_{\dot{1}5}}-\frac{\xi t_{1\dot{2}}t_{5\dot{1}}}{\mathcal{E}_{\dot{2}1}}
+
\frac{t_{1\dot{1}}t_{5\dot{2}}}{\mathcal{E}_{\dot{1}1}}-\frac{\xi t_{1\dot{1}}t_{5\dot{2}}}{\mathcal{E}_{\dot{2}5}}\Big)
-\Lambda_{\mathbf{r}_3j'}^{\alpha_5\alpha_3}\delta_{\mathbf{r}_5\mathbf{r}_3} \Big(
\frac{t_{1\dot{2}}t_{5\dot{1}}}{\mathcal{E}_{\dot{1}3}\mathcal{E}_{\dot{1}5}}
-
\frac{t_{1\dot{1}}t_{5\dot{2}}}{\mathcal{E}_{\dot{2}3}\mathcal{E}_{\dot{2}5}}
\Big)
\Big]
\Big\}.
\end{align}
where the sum of all indices are assumed in the above, except $\dot{\mathbf{r}}_{1,2}$.

\end{widetext}

\end{document}